# Second Harmonic Flexoelectricity

Zhiguo Wang [1,2,3,11], Yida Yang [4,11], Shuai Yang [5], Haoran Li [1], Chunchun Li [5], Zhibin Wen [1,2,3], Zhengqiu Xie [6], Wenbin Huang [7], Konstantin Shapovalov [8], Jiawang Hong [4], Longlong Shu [1,2,3]✉, Fei Li [5]✉, Gustau Catalan [9,10]✉

[1]. *School of Physics and Materials Science, Nanchang University, Nanchang 330031, Peoples R China.*

[2]. *Jiangxi Provincial Center for Interdisciplinary Studies in Future Display, Nanchang University, Nanchang 330031, Peoples R China.*

[3]. *The Jiangxi Provincial Engineering Technology Research Center for Advanced Functional Film Materials, Nanchang University, Nanchang 330031, Peoples R China.*

[4]. *Beijing Institute of Technology, Schools of Aerospace Engineering, Beijing 100081, Peoples R China*

[5]. *Xi'An Jiao Tong Univ, State Key Lab Mech Behav Mat, Educ Minist, Elect Mat Res Lab,Key Lab, Xian, Shaanxi, Peoples R China*

[6]. *College of Mechanical Engineering, Chongqing University of Technology, Chongqing 400050, Peoples R China*

[7]. *Chongqing University, State Key Laboratory of Mechanical Transmission, Chongqing 400044, Peoples R China*

[8]. *Universite de Liége, Theoretical Materials Physics, Q MAT, Sart Tilman Par Liege, Belgium*

[9]. *Institucio Catalana de Recerca i Estudis Avançats (ICREA), Barcelona, Catalonia.*

[10]. *Institut Catala de Nanociencia i Nanotecnologia (ICN2), CSIC–BIST, Campus Universitat Autonoma de Barcelona, Barcelona, Catalonia.*

[11]. *These authors contributed equally*: Zhiguo Wang, Yida Yang.

✉*e-mail*: llshu@ncu.edu.cn; ful5@xjtu.edu.cn; gustau.catalan@icn2.cat

## Abstract

Flexoelectricity (polarization induced by strain gradients) is assumed to be insensitive to space inversion in all materials: flexoelectric polarization should not depend on whether a crystal is bent upwards or downwards. In this work, we show that this assumption is wrong for ferroelectrics of all thickness and at all curvatures, owing to an intrinsic quadratic contribution that persists down to strain gradients of 1 $m^{-1}$ or less, which is millions of times smaller than conventional expectations for nonlinear flexoelectricity. As a consequence, ferroelectric crystals can generate second-harmonic voltage at twice the oscillation frequency of the bending.

Flexoelectricity (electric polarization in response to strain gradients) is a universal property of materials, having been observed in insulators [1,2], semiconductors [3-5], metals [6], and even bones [7] or ice [8,9]. The universality of flexoelectricity is guaranteed by the fourth rank nature of the flexoelectric tensor that couples strain gradient to polarization: being even-parity, it is compatible with materials of *any* symmetry [10]. Another consequence of even-parity is that flexoelectric polarization should be insensitive to space inversion in all materials, including those with that already have spontaneous polarization and thus a broken inversion symmetry. In other words: flipping a ferroelectric crystal upside down should have no effect on its flexoelectric coefficient.

Mathematics being an exact science, this expectation is of course mathematically correct. However, it rests on a physically incomplete picture -and is therefore ultimately wrong. As we will show, the bending-induced polarization of poled ferroelectrics cannot be adequately described by the linear flexoelectric tensor and includes a nonlinear term at all curvatures -we have observed nonlinearity at curvatures as small as $10^{-2}$ m$^{-1}$, the smallest we can measure in our laboratory, in stark contrast with the millions of times higher gradients (of the order of $10^{7}$ m$^{-1}$) previously thought to be required to reach a nonlinear regime [11[12][13]. As a consequence, the flexoelectric response of poled ferroelectrics becomes asymmetric with respect to their polar orientation. This unexpected result is shown in Figure 1 for poled ceramics of ferroelectric $BaTiO_3$ (BTO, Fig. 1).

To understand this effect, let us consider the simplest case of a monodomain ferroelectric beam with polarization normal to the surface (Fig. 1A). The beam, of thickness $h$ normal to the $z$ axis, is bent around the $y$ axis to produce a strain $\varepsilon_{xx} = Gz$, where $G$ is the curvature of the beam, and $-h/2 < z < h/2$ (Fig. 1B). The direction of the strain gradient is parallel or antiparallel to the ferroelectric polarization depending on whether the curvature is upward or downward, respectively. Being ferroelectric, the beam exhibits piezoelectricity in addition to flexoelectricity, and both effects can be included in a generalized Landau-based free energy:

$$F = F_P(P) + \frac{C}{2}\varepsilon^2 - qP^2\varepsilon - fP\frac{d\varepsilon}{dz} - \sigma\varepsilon \quad (1)$$

where $P = P_z$ is the out-of-plane polarization, $F_P(P)$ is the double-well potential (in the simplest case, $F(P) = \alpha P^2/2 + \beta P^4/4$ with $\alpha < 0$ and $\beta > 0$), $\varepsilon$ is the strain, $C$ is the elastic constant, $q$ is the electrostrictive coupling, $f$ is the flexo-coupling coefficient, and $\sigma = CGz$ is the externally applied stress that produces elastic in-plane strain $\varepsilon = \sigma/C = Gz$ and strain gradient $\nabla\varepsilon = G$. A self-consistent minimization of (1) results in the following polar response to bending, taken as the average across the crystal up to the second power of the induced strain gradient $G$ (see Methods I):

$$\Delta P_{\mathrm{eff}} = \mu_{\mathrm{eff}}G + w_{\mathrm{eff}}G^2 \quad (2)$$

where $\mu_{\mathrm{eff}}$ is the conventional (linear) flexoelectric coefficient and $w_{\mathrm{eff}}$ is the quadratic one. The analytically calculated flexoelectric coefficients (Methods I) are:

$$\mu_{eff} = \chi f \quad \text{and} \quad w_{eff} = -\chi^2\left(\frac{3f^2}{2P_0} + \frac{q^2P_0h^2}{6}\right) = -\frac{3\chi^2 f^2}{2P_0} - Ah^2 \quad (3)$$

Where $\chi$ is the dielectric susceptibility, f is the flexocoupling coefficient, and $A = \frac{q^2\chi^2P_0}{6} = \frac{C^2d^2}{24P_0}$ is a new term that we label as *quadratic flexo-thickness coefficient*. The polarization as a function of strain gradient is thus not a straight line but a parabola (Fig. 1C).

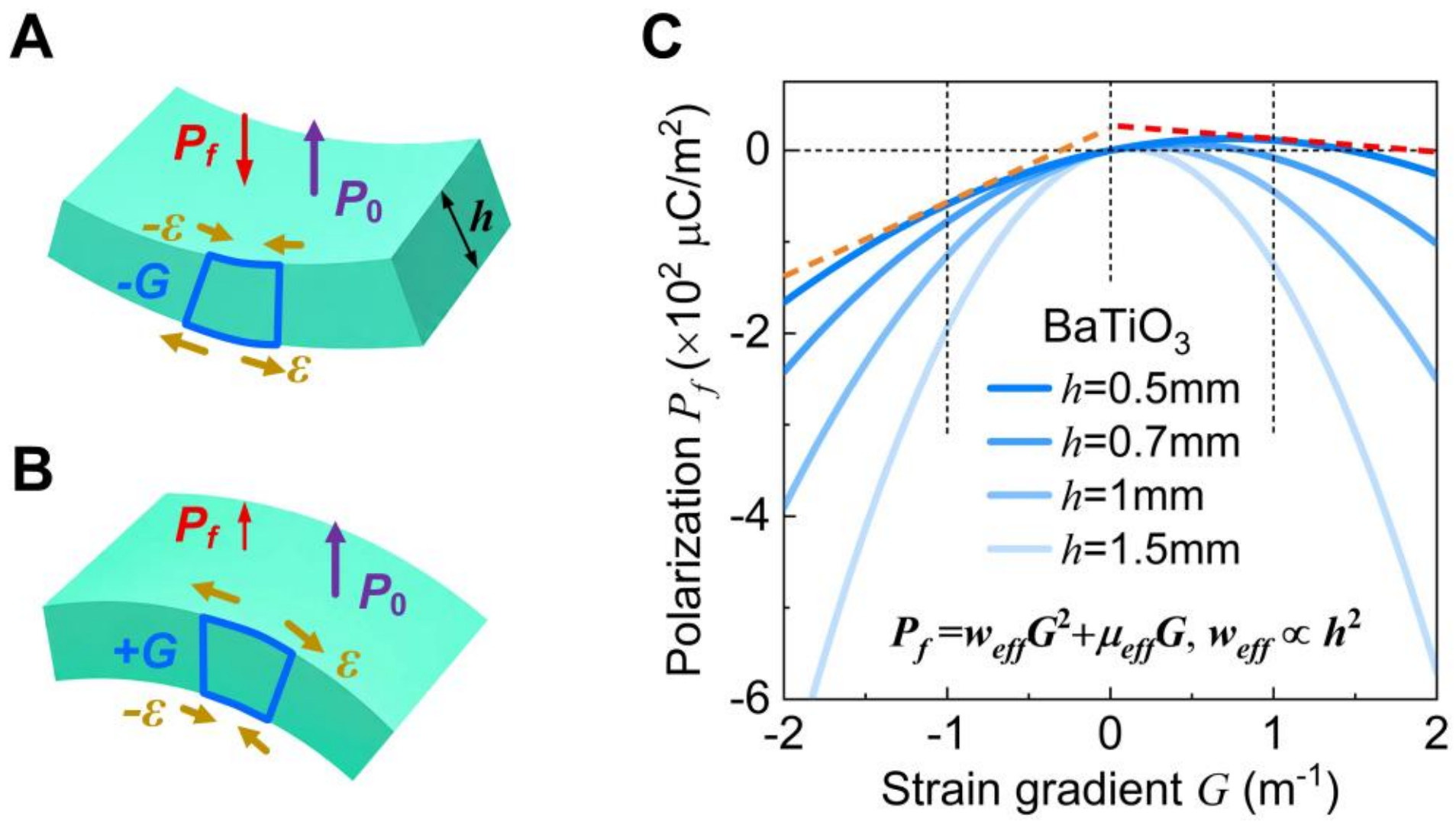

**Figure 1**. Flexoelectric polarization schematic (**A, B**) and results showing asymmetric flexoelectricity, with the magnitude and sign of the flexoelectric polarization depending on the relative orientation of the strain gradient with respect to the ferroelectric polarization. (**C**) Bending-induced polarization as a function of strain gradient G for poled ceramics of ferroelectric $BaTiO_3$ of different thickness. The curves are constructed from the values of the linear ($\mu_{eff}$) and quadratic ($w_{eff}$) coefficients measured by harmonic analysis of the charge generated by oscillatory bending. The result are thickness-dependent instead of the thickness-independent straight line expected from standard linear flexoelectricity, and both the magnitude and sign of the average polarization as a function of strain gradient (dashed orange and red lines) depend on whether the strain gradient is positive or negative, meaning that the total effective flexoelectric coefficient may be positive, negative or even zero depending on the magnitude and sign of the strain gradient with respect to the ferroelectric polarity.

The nonlinear flexoelectric coefficient has not one but two contributions. The first, proportional to $f^2$, is essentially a nonlinear susceptibility: it represents the higher-order response of the ferroelectric polarization to the flexoelectric field $E_{fl} \propto fG$. As such, it is a perturbative effect that requires large strain gradients and may be seen as the analytical embodiment of models based on second-order corrections [12],[13]. Its key feature is its explicit odd-parity dependence on $P_0$, which implies that the sign changes if we invert the polarization.

The second term (flexo-thickness coefficient) is related to surface piezoelectricity: as we bend the crystal, the tensile strain on the convex side reduces the out-of-plane ferroelectric polarization, thus decreasing the local piezoelectric coefficient, which is proportional to $P$. The reverse happens at the concave side: local compressive strain *increases* polarization and so too the local piezoelectric coefficient. Thus, as we bend the crystal, the initially-homogeneous piezoelectric becomes inhomogeneous and starts acting as a bimorph actuator [14]. Since surface strain grows with thickness and the piezoelectric coefficients also change with thickness, the surface polarization induced by bending changes as thickness squared. It is this thickness-squared dependence that makes the effect dominant at macroscopic scales. It is also worth noting that the nonlinear piezoelectricity that accounts for this effect is unrelated to high-order coefficients [16]; the value of $A$ is purely given by first-order parameters of the ferroelectric phase: the elastic constant $C$, the linear piezoelectric coefficient $d$, and the spontaneous polarization $P_0$. We also note that this term is susceptible to depolarizing effects, which we have neglected to simplify the analysis. Such simplification will not alter qualitatively the results but may reduce them quantitatively. We nevertheless find that the depolarization-free expression for A (Methods I) reproduces well the experimental results, thus validating the simplification.

One consequence of Eq. 2 is that the effective flexoelectric coefficient (derivative of the polarization with respect to the strain gradient) must become zero when $G' = -\mu_{\mathrm{eff}}/(2w_{\mathrm{eff}})$. Therefore, an empirically measured flexoelectric coefficient may be positive, negative, or zero, depending on thickness, polarity, and bending magnitude, which can account for enormous discrepancies between measurements of flexoelectricity for any given ferroelectric. We can also use this prediction of a "neutral gradient" $G'$ to test the accuracy of the model. First, we note that $G'$ is different for very thin and very thick samples. According to Eq. 3, in the thin limit $h \ll 3f/(qP_0)$, quadratic flexoelectricity is determined by the first (thickness-independent) flexocoupling term, while in the thick limit $h \gg 3f/(qP_0)$ it is dominated by quadratic flexo-thickness. The threshold value is $h_t \equiv 3f/(qP_0) \sim 1$ to $10\ \mu$m, using standard values for BTO (Methods II).

According to Eq. 3, the total bending-induced polarization of BTO thin films ($h \ll h_t$) will be zero when

$$G' = -\mu_{\mathrm{eff}}/(2w_{\mathrm{eff}}) = P_0/(3\chi f) \sim 10^5 \sim 10^6\ \mathrm{m^{-1}} \tag{4}$$

Such strain gradients are equivalent to curvature radii of ~1-10 microns, quite common in wrinkled oxide membranes [17],[18]. Thus, even if large flexoelectric effects may a priori be expected in ferroelectric membranes owing to their large achievable curvatures, nonlinear flexoelectricity can get in the way and frustrate such expectations.

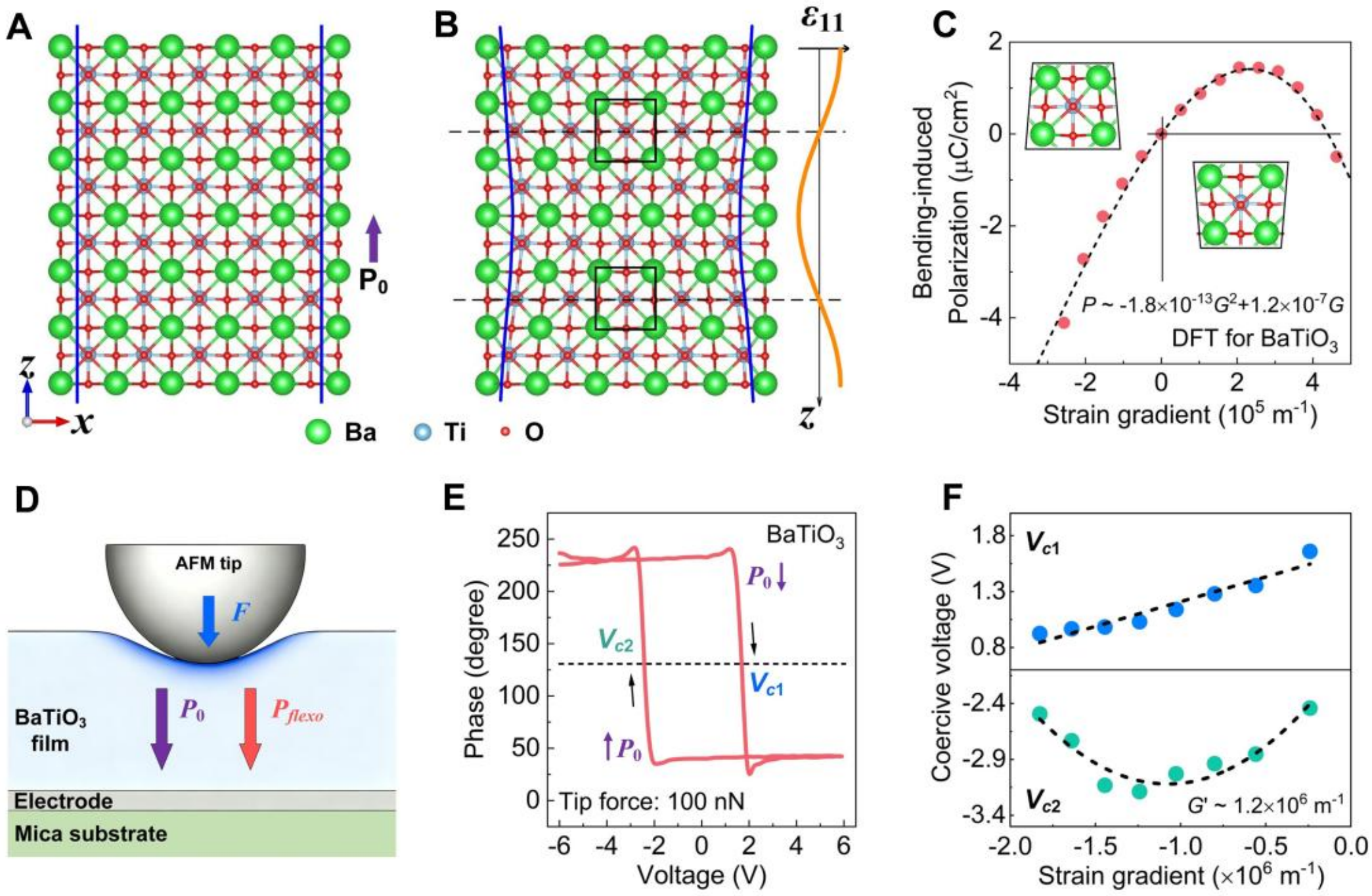

**Figure 2.** (**A, B, C**) **DFT of BTO**. (**A**) Initial atomic arrangement model, free from strain gradients. The supercell comprises 5×1×6-unit cells. The atoms in green, red, and silver represent barium, oxygen, and titanium, respectively. (**B**) Atomic arrangement emulating the strain profile depicted in the right-hand-side inset. The blue lines enclose the region within which the tensile strain is applied. The unit cells selected for the calculation of flexoelectric polarization in the $z$ direction are highlighted with a black trapezoidal frame where, by geometry, the local strain is zero and the strain gradient is maximal. (**C**) Polarization as a function of strain gradient calculated from (**B**). (**D**) Scheme of an experiment where stress is applied to the $BaTiO_3$ film using an AFM tip to induce a strain gradient. (**E**) Polarization reversal measurements are conducted using PFM, and the coercive voltage corresponds to the switching barrier, which is given by the sum of the bulk coercive field and the flexoelectric field. (**F**) Measured coercive voltages as functions of strain gradient. The transition voltage $V_{c1}$ continuously decreases, while $|V_{c2}|$ first increases and then decreases, following a parabola whose minimum matches our expectation value.

We have verified the predicted critical curvature of films in two independent ways: a first-principles DFT simulation and actual measurements of BTO thin films. For DFT, we perform a finite-size simulation with a thickness of 6 unit-cells with closed-circuit boundary conditions to eliminate depolarization and initial polarization set as positive (upwards) (Fig. 2A). The strains of Ba ions were set as $\varepsilon_{xx} = \varepsilon_{\max} \cos(2\pi z / h)$, where $h$ is the thickness of the model. This arrangement produces strain gradient $\partial\varepsilon_{xx}/\partial z$ that is maximal in the regions highlighted with a square in Fig. 2B (see also Methods IV). We then allowed the remaining ions to relax and, based on their displacements in the highlighted regions, we extracted the induced local polarization as a function of strain gradient $G = 2\pi\varepsilon_{\max}/h$ (Fig. 2C). The tangent slope at $G = 0$ yields the linear coefficient $\mu_{eff}$ = 0.12 μC/m, in good agreement with prior DFT calculations [19] (experimental values are higher due to additional piezoelectric contributions [20]). Crucially, the results confirm the parabolic dependence of P on strain gradient, with a maximum at $G' = -\mu_{\text{eff}}/2w_{\text{eff}} \sim 3 \times 10^5\ m^{-1}$, in excellent agreement with our predicted range.

For the experimental verification, we prepared $BaTiO_3$(111)/$SrRuO_3$/$SrTiO_3$/mica (BTO/SRO/STO/mica, see Methods V). Since film and substrate are bonded only by weak van der Waals forces, the film can slide over the mica and thus the film can bend quite freely [21]. The BTO film was about 8 nm-thick and of good epitaxial quality (Supplementary Material S3-S5). Stress was applied to the BTO film using an atomic force microscope (AFM) probe (Fig. 2D) and we used COMSOL to correlate the tip force with the strain gradient maximum under the tip, simplified as a sphere (Fig. S6). For BTO with (111) orientation out-of-plane, the linear flexoelectric coefficient is negative ($\mu_{eff} < 0$) [22]. By analyzing the switching hysteresis of the piezoresponse as a function of tip force (Fig. 2E shows the typical curve for 100 nN, other forces are in Supplementary Material S7), we track the coercive voltage as a function of strain gradient. As shown in Fig. 2F, the positive coercive voltage decreases with increasing pressure, as expected – a large enough strain gradient can switch polarization even without applied voltage [23]. The negative one, however, initially increases but then decreases; the turning point marks the gradient $G$' at which linear flexoelectricity is compensated by quadratic, and the experimental value $G' \sim 1.2 \times 10^6\ m^{-1}$ is within the expectation range given by Eq. 4.

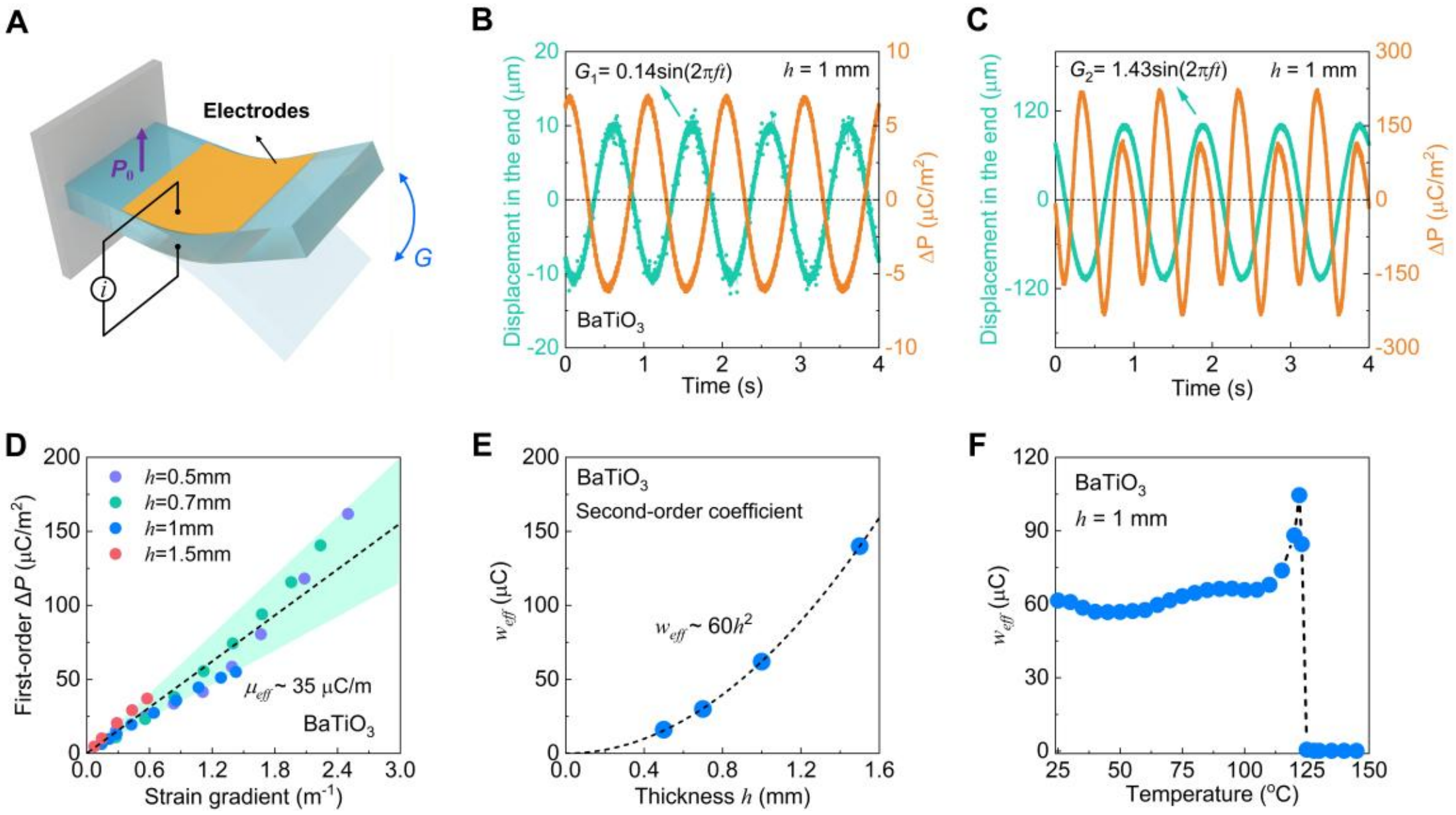

**Figure 3**. **First and second harmonic flexoelectricity in BTO**. (**A**) Schematic of the experiment, whereby an alternate strain gradient is applied to a pre-poled sample (Supplementary Material S7 for actual measurements). (**B**) first harmonic (linear) flexoelectric response at low gradients (0.14 $m^{-1}$), (**C**) flexoelectric response at bigger applied strain gradients (1.4 $m^{-1}$) with a significant second-harmonic component, (**D**) dependence of the linear flexoelectric polarization on strain gradient for samples of different thickness: the flexoelectric coefficient (slope of the linear least-squares fit to the data) is independent of thickness, as expected; (**E**) dependence of the quadratic flexoelectric coefficient on sample thickness, showing the predicted parabolic dependence (**F**) temperature dependence of the quadratic flexoelectric coefficient (heating rate = 3 K/min), vanishing above $T_c$.

A consequence of quadratic flexoelectricity is that an oscillatory deformation can now generate a second-harmonic signal, that is, an alternate current (ac) output at twice the oscillation frequency of the cantilever. We have verified this prediction in a set of poled ceramic $BaTiO_3$ samples ($h$ = 0.5, 0.7, 1 and 1.5 mm; Fig. 3A is a sketch of the experiment). Being much thicker than $h_t$, the nonlinear flexoelectric coefficient (Eq. 3) can be simplified as $w_{eff} = -P_0\,(\chi q h)^2/6 = -Ah^2$, and the bulk critical curvature is $G' = \chi f/(Ah^2) = 3f/(q^2 \chi P_0 h^2)$. By substituting the relevant physical parameters of BTO into the equation, we get $A$ ~ 50 C/m², and $G$'=0.28 $m^{-1}$ (results for $h$ = 1mm, all other samples are in Supplementary Material S9). The results indicate that, for low strain gradients ($G < G$') the linear (first harmonic) signal dominates (Fig. 3B), but bigger strain gradients ($G$ ~ $G$') generate a comparable electrical signal at double the vibration frequency (Fig. 3C). From the Fourier transform of the oscillatory polarization, we extract the first and second order amplitudes (see Fig. S10). The first-harmonic as a function of strain gradient is plotted in Fig. 3D and shows the same trend for all the samples, with $\mu_{eff} \sim 35$ µC/m. Meanwhile, the second harmonic (Fig. 3E, see also Fig. S11 and values in Table S1) is proportional to $t^2$. Fitting the parabola with Eq. 3, we obtain $A \approx 60\ C/m^2$, in excellent agreement with the theoretical prediction.

The flexoelectric coefficient is thus never fully linear: neither in the thin film regime, nor in thick samples. Interestingly, while one might have expected nonlinearity to be more prominent for very thin samples (where larger gradients are easier to achieve), the contribution from surface piezoelectricity is such that the opposite is true: thicker samples show a stronger second-harmonic effect at much smaller gradients. This quadratic effect only exists because of ferroelectricity: the dependence on temperature (Fig. 3F) shows that, above the Curie temperature of BTO, $w_{eff}$ vanishes (in Supplementary Material S12 we provide the output waveforms of BTO at 122 ºC and 130 ºC). We have tested the generality of these results by exploring other materials: relaxor ferroelectric PMN-PT *ceramic*, and PMN-PT *single crystal* (Supplementary Materials S13 and S14). Both showed the same trends, so quadratic flexoelectricity does not arise from grain boundaries [24] nor require specific compositions.

The similarity between the magnitudes of the first and second harmonic signals can be exploited to make a device which, for the same oscillatory input, selectively generates equivalent voltage outputs at the first or the second harmonic. We achieve such harmonic distillation by topological engineering of electrode connectivity and ferroelectric poling in the scheme shown in Fig. 4A, where two identical sets of parallel-plate electrodes are deposited in the same ferroelectric slab and labelled as a-b (capacitor 1) and c-d (capacitor 2).

If the a and c, b and d electrodes are joined (Abbreviation: ac-bd), the two capacitors are in parallel (Fig. 4B) and the total charge induced by bending is the sum of their flexoelectric contributions (Fig. S15). Because the nonlinear term has an odd dependence on polarity, the antiparallel nonlinear components mutually cancel and the measured polarization (charge divided by the area of the capacitors) is just $P_f = \mu G$ (the analysis is shown in Methods VIII). Being linear with the gradient, the output is purely first harmonic. We confirm this prediction experimentally (Fig. 4C) using a BTO ceramic slab with the electrode configuration depicted in Fig. 4A (the individual output of each pair of electrodes is shown in Fig. S16). The observation of purely linear flexoelectricity in capacitors with antiparallel polarization explains why the relationship between flexoelectric polarization and strain gradient *appears* to be linear in measurements of *unpoled* ferroelectrics [20][22]: random (unpoled) domain orientations act de facto as capacitors of opposite polarization.

In contrast, if the a and d, b and c electrodes are joined, the electrode configuration is topologically equivalent to bending two capacitors of identical polarity in opposite directions (Fig. 4D), which cancels the linear flexoelectric components because the gradients point in opposite directions, while at the same time adds the term that depends on the square of the gradient (Fig. S15B), so $P_f = wG^2$ (further analysis is shown in Methods VIII). The output is thus a perfect second-harmonic (Fig. 4E). Such bi-harmonic generation was replicated in PMN-PT (Supplementary Material S17).

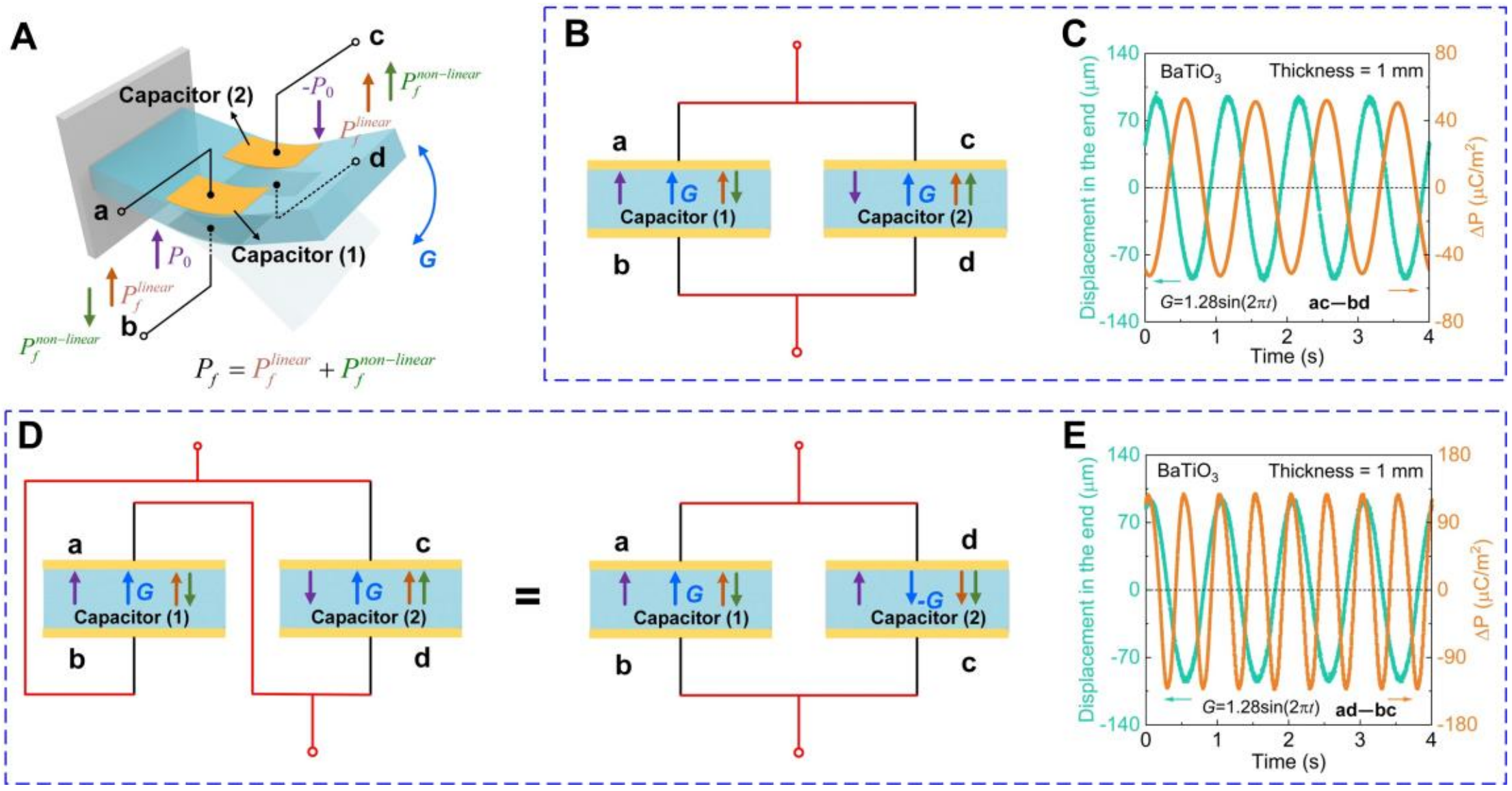

**Figure 4**. **Harmonic distillation through connection topology**. (**A**) Diagram of the initial polarization and of the electrodes positioned on a ferroelectric slab, with two capacitors (electrodes

a-b and c-d) poled in opposite directions. (**B**) Connecting the electrodes ac and bd results in equivalent parallel capacitors where (**C**) the flexoelectric output signal has the same frequency as the displacement (pure first harmonic). (**D**) Connecting the electrodes ad-bc is equivalent to flipping one of the capacitors around, resulting in (**E**) flexoelectric output at double the frequency of the displacement (pure second harmonic).

The frequency-doubling output is somewhat reminiscent of electrostriction, but in reverse: while electrostriction generates a deformation proportional to the square of voltage [26], second harmonic flexoelectricity generates a voltage perfectly proportional to the square of deformation, even under very small deformations. The results thus fill a functional void, in addition to upending (literally) the relationship between flexoelectricity and symmetry in ferroelectrics.

**ACKNOWLEDGMENT**

This work was supported by the National Natural Science Foundation of China (Nos. 52472124, 12404106, 52325205 and 12404101) and the Natural Science Foundation of Jiangxi Province (No. 20252BAC200609). ZG Wang and LL Shu acknowledge the support from Nanchang University. ICN2 is supported by the Severo Ochoa program from Spanish MCIN / AEI (Grant No. CEX2021-001214-S. GC acknowledges support from the EU through HORIZON-MSCA-SE-3D-TOPO action (Grant agreement ID: 101236483) and from Spain's MICINN through grant PID2023-148673NB-I00 (Project GRIPHO2).

## Methods

### I. Relationship between flexoelectric polarization and strain gradients from Landau theory

To obtain the response of polarization $P(z)$ and strain $\varepsilon(z)$ within the cantilever to an applied stress $\sigma(z)$ and/or electric field $E$, we minimize the Landau energy [Eq. (1)] with respect to $P(z)$ and $\varepsilon(z)$, using the following full expression for the energy:

$$F = F_P(P) + \frac{C}{2}\varepsilon^2 - qP^2\varepsilon - fP\frac{d\varepsilon}{dz} + \frac{\delta}{2}\left(\frac{dP}{dz}\right)^2 - \sigma\varepsilon - PE,$$

where we have added the polarization correlation term controlled by $\delta$, and the explicit coupling between polarization and electric field $-PE$. To obtain the energy minimum, we solve the system of Euler-Lagrange equations:

$$\frac{\partial F}{\partial P} - \frac{d}{dz}\frac{\partial F}{\partial(dP/dz)} = \frac{dF_P}{dP} - \delta\frac{d^2P}{dz^2} - 2qP\varepsilon - f\frac{d\varepsilon}{dz} - E = 0, \quad \text{(M1.1)}$$

$$\frac{\partial F}{\partial \varepsilon} - \frac{d}{dz}\frac{\partial F}{\partial(d\varepsilon/dz)} = C\varepsilon - qP^2 + f\frac{dP}{dz} - \sigma = 0. \quad \text{(M1.2)}$$

Solving the equation for the strain (M1.2) results in

$$\varepsilon = \frac{q}{C}P^2 - \frac{f}{C}\frac{dP}{dz} + \frac{\sigma}{C}. \tag{M1.3}$$

As discussed in the main text, in the following we consider the specific case $\sigma(z) = CGz$, $-\frac{h}{2} < z < \frac{h}{2}$, which describes the stresses produced by bending of a cantilever of thickness $h$. The third term in Eq. (M1.3), describing the elastic response to the applied stress, thus naturally gives contribution to strain $\varepsilon_{\mathrm{elast}} = Gz$ and to strain gradient $\frac{d\varepsilon_{\mathrm{elast}}}{dz} = G$.

Substituting Eq. (M1.3) into the Euler-Lagrange equation for polarization (M1.1) then results in the following:

$$\frac{\partial}{\partial P}\left(F_P(P) - qGzP^2 - \frac{q^2}{2C}P^4\right) - \left(\delta - \frac{f^2}{C}\right)\frac{d^2P}{dz^2} - fG - E = 0,$$

from where it is apparent that the strain gradient leads to a $z$-dependent renormalization of the $P^2$ coefficient in the double-well potential expression through the electrostrictive coupling, producing an asymmetric polar response to the strain gradient at the top and at the bottom surfaces of the cantilever. For the following use, we rewrite the equation for polarization as follows:

$$\frac{\partial}{\partial P}\left(F_P^*(P) - qGzP^2\right) - fG - E = 0, \tag{M1.4}$$

$$F_P^*(P) = F_P(P) - \frac{q^2}{2C}P^4,$$

and we neglect the effect of the term $\left(\delta - \frac{f^2}{C}\right)\frac{d^2P}{dz^2}$ under the assumption that the thickness of the cantilever $h$ far exceeds the characteristic scale $\xi \sim \left(\left(\delta - \frac{f^2}{C}\right)\chi\right)^{\frac{1}{2}}$ of variations of polarization (typically $\xi \sim 1$ nm in ferroelectrics).

In the absence of strain gradient and electric field, the spontaneous polarization is described by setting $G = 0$, $E = 0$ in Eq. (M1.4): $\left.\frac{dF^*}{dP}\right|_{P=P_0} = 0$. For instance, the classic double-well potential $F^*(P) = \frac{\alpha}{2}P^2 + \frac{\beta}{4}P^4$ leads to the equation for spontaneous polarization $\alpha P_0 + \beta P_0^3 = 0$ and $P_0 = \sqrt{-\frac{\alpha}{\beta}}$.

In the following, we will write the linear and quadratic responses to the strain gradients in terms of the linear and quadratic permittivity, as well as the linear piezoelectric coupling constants. Let us first list the explicit expressions for the three latter parameters. At first, we describe the effect of applying a uniform electric field $E$ on polarization, $P(E) = P_0 + \chi E + \kappa E^2$, in a stress-free sample ($G = 0$). Equation (M1.4) then rewrites as:

$$\frac{dF_P^*}{dP} - E = 0. \tag{M1.5}$$

The linear and quadratic responses to the applied electric field $\chi = \left.\frac{dP}{dE}\right|_{E=0}$ and $\kappa = \left.\frac{1}{2}\frac{d^2P}{dE^2}\right|_{E=0}$ can be obtained by differentiating Eq. (M1.5) with respect to $E$:

$$\frac{d}{dE}\left(\frac{dF_P^*}{dP} - E\right) = \frac{d^2F_P^*}{dP^2}\frac{dP}{dE} - 1 = 0,$$

$$\chi = \left.\frac{dP}{dE}\right|_{E=0} = \left.\left(\frac{d^2F_P^*}{dP^2}\right)^{-1}\right|_{E=0},$$

$$\frac{d^2}{dE^2}\left(\frac{dF_P^*}{dP} - E\right) = \frac{d^3F_P^*}{dP^2}\left(\frac{dP}{dE}\right)^2 + \frac{d^2F_P^*}{dP^2}\frac{d^2P}{dE^2} = 0,$$

$$\kappa = \left.\frac{1}{2}\frac{d^2P}{dE^2}\right|_{E=0} = -\left.\frac{1}{2}\frac{d^3F_P^*}{dP^3}\left(\frac{dP}{dE}\right)^2\left(\frac{d^2F_P^*}{dP^2}\right)^{-1}\right|_{E=0} = -\left.\frac{\chi^3}{2}\frac{d^3F_P^*}{dP^3}\right|_{E=0}.$$

For example, using the classic double-well potential $F^*(P) = \frac{\alpha}{2}P^2 + \frac{\beta}{4}P^4$ leads to $\chi = -\frac{1}{2\alpha} > 0$ and $\kappa = -\frac{3\chi^2}{2P_0} < 0$ at $P = +P_0$.

Next, we consider linear and quadratic piezoelectricity, i.e. the uniform responses of polarization to an applied uniform stress $\sigma$ in the absence of electric field. Let us write polarization as $P(\sigma) = P_0 + d\sigma + b\sigma^2$, where $d$ and $b$ are linear and quadratic direct piezoelectric coefficients (while we do not use $b$ in the main text, it is instrumental to obtain the quadratic piezoelectricity explicitly). The total strain produced by an applied stress is given by Eq. (M1.3), $\varepsilon = \frac{q}{C}P^2 + \frac{\sigma}{C}$. Then linear and quadratic piezoelectric coefficients $d = \left.\frac{dP}{d\sigma}\right|_{\sigma=0}$ and $b = \left.\frac{1}{2}\frac{d^2P}{d\sigma^2}\right|_{\sigma=0}$ can be found by substituting this expression for strain into Eq. (M1.1),

$$\frac{dF_P}{dP} - \frac{2q^2P^3}{C} - \frac{2qP\sigma}{C} = \frac{dF_P^*}{dP} - \frac{2qP\sigma}{C} = 0,$$

and differentiating it with respect to $\sigma$:

$$\left[\frac{d}{d\sigma}\left(\frac{dF_P^*}{dP} - \frac{2qP\sigma}{C}\right)\right]_{\sigma=0} = \left[\left(\frac{d^2F_P^*}{dP^2} - \frac{2q\sigma}{C}\right)\frac{dP}{d\sigma} - \frac{2qP}{C}\right]_{\sigma=0} = \frac{d}{\chi} - \frac{2qP_0}{C} = 0,$$

$$d = \frac{2q\chi P_0}{C},$$

$$\left[\frac{d^2}{d\sigma^2}\left(\frac{dF_P^*}{dP} - \frac{2qP\sigma}{C}\right)\right]_{\sigma=0} = \left[\frac{d^3F_P^*}{dP^3}\left(\frac{dP}{d\sigma}\right)^2 - \frac{4q}{C}\frac{dP}{d\sigma} + \left(\frac{d^2F_P^*}{dP^2} - \frac{2q\sigma}{C}\right)\frac{d^2P}{d\sigma^2}\right]_{\sigma=0}$$

$$= -\frac{2\kappa}{\chi^3}d^2 - \frac{2d^2}{\chi P_0} + \frac{2b}{\chi} = 0,$$

$$b = \frac{d^2}{P_0}\left(1 + \frac{\kappa P_0}{\chi^2}\right),$$

$$P(\sigma) = P_0 + d\sigma + \frac{d^2}{P_0}\left(1 + \frac{\kappa P_0}{\chi^2}\right)\sigma^2,$$

where we have expressed electrostriction $q$ in terms of the piezoelectric constant $d$, $q = \frac{Cd}{2\chi P_0}$, and used $\frac{d^2F_P^*}{dP^2} = \frac{1}{\chi}$, $\frac{d^3F_P^*}{dP^3} = -\frac{2\kappa}{\chi^3}$.

Finally, we describe the response of polarization to the strain gradients in the absence of electric fields. Writing polarization as $P(G) = P_0 + \mu G + wG^2$, the linear and quadratic responses to the applied strain gradient $\mu = \left.\frac{dP}{dG}\right|_{G=0}$ and $w = \left.\frac{1}{2}\frac{d^2P}{dG^2}\right|_{G=0}$ can be obtained by differentiating Eq. (M1.4) with respect to $G$:

$$\left[\frac{d}{dG}\left(\frac{dF_P^*}{dP} - 2qGzP - fG\right)\right]_{G=0} = \left[\left(\frac{d^2F_P^*}{dP^2} - 2qGz\right)\frac{dP}{dG} - 2qzP - f\right]_{G=0} = \frac{\mu}{\chi} - \frac{Cdz}{\chi} - f = 0,$$

$$\mu = \chi f + Cdz,$$

$$\left[\frac{d^2}{dG^2}\left(\frac{dF_P^*}{dP} - 2qGzP - fG\right)\right]_{G=0} = \left[\frac{d^3F_P^*}{dP^3}\left(\frac{dP}{dG}\right)^2 - 4qz\frac{dP}{dG} + \left(\frac{d^2F_P^*}{dP^2} - 2qGz\right)\frac{d^2P}{dG^2}\right]_{G=0}$$

$$= -\frac{2\kappa}{\chi^3}\mu^2 - \frac{2Cdz}{\chi P_0}\mu + \frac{2w}{\chi} = 0,$$

$$w = \frac{Cdz}{P_0}\mu + \frac{\kappa}{\chi^2}\mu^2 = \frac{Cdz}{P_0}(\chi f + Cdz) + \frac{\kappa}{\chi^2}(\chi f + Cdz)^2,$$

$$P(G) = P_0 + (\chi f + Cdz)G + \left[\frac{Cdz}{P_0}(\chi f + Cdz) + \frac{\kappa}{\chi^2}(\chi f + Cdz)^2\right]G^2.$$

We estimate the experimentally extracted polarization by the average value within the cantilever:

$$\Delta P_{\text{eff}} = \frac{1}{h}\int_{-\frac{h}{2}}^{\frac{h}{2}} P(z)dz - P_0 = \mu_{\text{eff}}G + w_{\text{eff}}G^2,$$

$$\mu_{\text{eff}} = \chi f, \qquad w_{\text{eff}} = \kappa f^2 + \frac{C^2d^2h^2}{12P_0}\left(1 + \frac{\kappa P_0}{\chi^2}\right),$$

obtaining Eq. (2) of the main text.

Finally, taking a simple double-well potential $F_P^*(P) = \frac{\alpha P^2}{2} + \frac{\beta P^4}{4}$ with $\kappa = -\frac{3\chi^2}{2P_0}$, we obtain Eq. (3) of the main text.

## II. Sample preparation and characterization

### BaTiO3 and Sm-PMN-30PT ceramics

$BaTiO_3$ ceramics were prepared by solid-phase sintering. $BaCO_3$ (99.99%, Aladdin), $TiO_2$ (99.99%, Aladdin) powders as the raw materials by molar ratio matching. BTO powders are formed by pre-firing at 1000 °C for 3h. These powders were granulated with polyvinyl alcohol solution and pressed into shape at a pressure of 20 MPa. Then, the BTO ceramics were sintered at 1250 °C for 6h in Muffle furnace. The samples were cut into pieces with thicknesses of 0.5, 0.7, 1, and 1.5 mm using a diamond cutting machine after the samples were sintered. They were then annealed through a tube furnace at 800 °C for 20 h in an air atmosphere to eliminate oxygen vacancies caused by sintering. The measured coefficients for $BaTiO_3$ ceramics (Supplementary Material S2) were $P_0$=10 μC/cm$^2$, $Q=q/C$=0.0425 m$^4$ C$^{-2}$, and $C$=150 GPa. The flexocoupling coefficient is $f$~1 - 10 V [20]).

Sm-PMN-30PT ceramics were prepared using a simple mixed oxide method, while a precursor method was used to fabricate the $MgNb_2O_6$ precursor powders [27]. Specifically, samples were calcined at 850 °C for 2h and sintered at 1230 °C for 2h. Thermal annealing was conducted at 900 °C for 10h to eliminate residual PbO on the surface of the ceramics. The sample was cut into long strips with a thickness of 0.7 mm.

PMN-PT (100) single crystals were commercially acquired (TOPVENDOR, Beijing, China) and double-side polished. All the samples were applied an electric field to make them generate an initial upward polarization, for the measurement of flexoelectric performance.

### $BaTiO_3/SrRuO_3/SrTiO_3$/mica films

High-quality $BaTiO_3$ (BTO) thin films were grown on mica (001) substrates by pulsed laser deposition. $SrTiO_3$ (STO) was first deposited on the substrate surface as a buffer layer, followed by the deposition of $SrRuO_3$ (SRO) as the bottom electrode. During the deposition of the STO buffer layer and SRO electrode layer, the substrate temperatures were set to 700°C and 680°C, respectively. The oxygen partial pressure was maintained at 100 mTorr, while the laser fluence and repetition rate were 1.38 mJ/cm$^2$ and 10 Hz, respectively. During the deposition of the BTO thin films, the substrate temperature was maintained at 750 °C, the oxygen partial pressure was kept at 10 mTorr, and the laser fluence and repetition rate were 1.38 mJ/cm$^2$ and 8 Hz, respectively.

## III. Characterization for macroscopic materials

The functional parameters of materials (include the coefficients of Eq. 2) were measured as follows:

**XRD**: The phase composition and crystallographic orientation of our cells was characterized by XRD using a D/Max-IIIC X-ray diffractometer (Rigaku, Tokyo, Japan) with Cu Ka radiation to understand the crystallization of the perovskite.

**SEM**: The microstructures of the natural surfaces and acid-etched surfaces were observed by a scanning electron microscopy (SEM, Model JSM 6710F, Jeol, Japan).

**Dielectric**: The dielectric properties of the ceramics were measured by a broadband dielectric spectrometer ((DMS3000, Partulab, Germany)) and an impedance analyzer (E4294A, Agilent, Palo Alto, USA).

**Ferroelectricity**: The ferroelectric hysteresis loops (P-E loops) were analyzed using a ferroelectric integrated test system (TF Analyzer 3000, aixACCT, Germany).

**Electrostriction**: The electrostrictive performances were analyzed using the ferroelectric integrated test system as well, the test temperature is 200 °C to ensure that the sample is in the paramagnetic phase.

**Elastic constant**: Force curve measurements on $BaTiO_3$ ceramics were performed using a Cypher ES atomic force microscope (Asylum Research, USA) operating in contact mode under ambient conditions. A CDT-NCLR conductive diamond probe (Nanosensors, Switzerland) was employed for all measurements, nominal spring constant: 72 N/m, resonant frequency: ~210 kHz, tip radius: ~100 nm) were employed to ensure electrical conductivity and mechanical robustness. To ensure representative data, force curves were acquired at 10 randomly selected regions across the polished surface of $BaTiO_3$ ceramic samples, with 5 individual curves recorded per region. All force curve data were processed using Asylum Research's Igor Pro-based software, and extraction of key mechanical parameters such as Young's modulus.

## IV. DFT for $BaTiO_3$

To further deepen our understanding of the experimental findings, we performed first-principles calculations to investigate the relationship between polarization and strain gradient in bulk $BaTiO_3$ (BTO). The whole calculations were conducted using the Perdew-Burke-Ernzerhof (PBE) with Projector-Augmented Wave (PAW) potentials, as implemented in the Vienna *ab initio* simulation package (VASP) [28]. The Monkhorst-Pack scheme employed for k-point grid sampling in the irreducible Brillouin zone was set to 4×2×1. A plane-wave cutoff energy of 500 eV was employed. Atomic positions were relaxed using the conjugate gradient algorithm until the residual forces were smaller than 0.005 eV/Å. We adopted the computational model proposed by Xu's approach [29], allowing for eliminating polarization contributions arising from non-equivalent surface terminations.

Fig. 2 shows the established configuration with 5×1×6 unit cells, with its energy fully relaxed by conjugate gradient algorithm. Periodic boundary condition is applied in all direction. There is only one unit cell in the y direction in the supercell. Strain gradient is created by displacing every atom along $x$ axis, according to a cosine displacement function $\varepsilon_{11} = \varepsilon_{\max} \cos\left(2\pi z / h\right)$, $\varepsilon_{\max}$ being the maximum strain and $h$ standing for the thickness of the model (see Fig. 2A), after that, the energy of the whole structure is fully minimized, while maintaining the Ba atom in a fixed position. Since the experimentally measured polarization includes both piezoelectric and flexoelectric contributions, we selected the unit cell where both strain and strain gradient are nonzero (highlighted by the trapezoidal frame in Fig. 1A-C) to evaluate the relationship between polarization and strain gradient. With the derivative of $\varepsilon_{11}$ with respect to $z$, the magnitude of the strain gradient $\frac{\partial \varepsilon_{11}}{\partial x_3}$ is computed to be ~$10^7$ $m^{-1}$, attainable in experimental measurements. Born effective charges are evaluated by using a density-functional perturbation theory (DFPT) [30]. With them, the polarization can be defined as: $P = \frac{e}{\Omega_c} \sum_j w_j Z_j^* u_j$, where $\Omega_c$ is the volume of the unit cell considered, $u_j$ denotes the atomic coordinates in the $z$ direction in the unit cell. Index $j$ spans all atoms contained in the unit cell. $Z_j^*$ is the Born effective charges of the bulk BTO. $w$ means weights that are set to 1/8 for Ba, 1 for Ti, and 1/2 for O.

## V. Characterization for $BaTiO_3/SrRuO_3/SrTiO_3$/mica

**Structural characterization:** X-ray diffraction measurements, including *2θ–ω* scans, *φ* scans, reciprocal space mapping (RSM), and X-ray reflectivity (XRR), were performed using a high-resolution X-ray diffractometer (SmartLab, Rigaku, JPN) equipped with a Cu Kα radiation source ($\lambda$ = 1.54 Å). The *2θ–ω* scans were used to determine the out-of-plane lattice parameters of the films and membranes, while *φ* scans were conducted to examine the in-plane epitaxial relationship between the film and substrate. RSM measurements around selected asymmetric reflections were carried out to further evaluate the strain state, lattice relaxation, and crystallographic coherence of the films. In addition, XRR measurements were performed to assess the film thickness, density, and interfacial roughness.

**Ferroelectricity:** Piezoresponse force microscopy (PFM) measurements were performed using an atomic force microscope (Asylum Research Cypher ES, Oxford Instruments, UK) to investigate the local ferroelectric properties of the BTO thin films. Conductive ASYELEC-01-R2 probes with a Ti/Ir coating were used for the measurements.

**Strain gradients:** switching spectroscopy piezoelectric force microscopy (SS-PFM) measurements were performed on the BTO thin film. A fixed position on the film surface was selected, and different normal loads ranging from 100 nN to 1000 nN were applied to the AFM tip. ASYELEC-01-R2(2.8 N $m^{-1}$ in force constant) were used in the measurements, the corresponding pressure can be estimated with a tip radius of 25 nm (ASYELEC-01-R2) under the assumption that the tip-sample contact is represented by a disc with radius equivalent to the tip radius.

**PFM measurements for $BaTiO_3$ film:** under the 6 V DC bias condition, PFM phase hysteresis loops were recorded at each applied load to investigate the influence of mechanical force on the ferroelectric switching behavior.

## VI. The flexoelectric polarization *vs* strain gradient for BTO film

For BTO(111) film, under the action of the strain gradient, the total polarization can be expressed as,

$$\begin{cases} P_1 = P_0 + \mu_{eff} G + w_{eff} G^2, & w < 0 \\ P_2 = -P_0 + \mu_{eff} G - w_{eff} G^2, & w > 0 \end{cases} \quad \text{(M6.1)}$$

$P_1$ and $P_2$ respectively represent the total polarization when the ferroelectric

polarization is upward and downward, which is the result of the coupling of spontaneous and flexoelectric polarization. At this point, the quadratic coefficient originates from non-thickness term, $w_{eff} = -\frac{3\chi^2 f^2}{2P_0}$. As the applied voltage increases, when the polarization direction of the electric domain changes, it indicates that the external electric field will be equal to the electric field formed by the $P_1$ or and $P_2$.

(1) When the polarization is positive, the second-order tortuosity coefficient $w_{eff}$<0 and $G' = -\mu_{eff} / (2w_{eff}) < 0$. Therefore, when the tip is pressed to apply a negative strain gradient, the hysteresis electric field $V_{d1}$ continues to decrease, the corresponding absolute value of total polarization intensity $P_1$ also decreases continuously accordingly.

(2) When the polarization is negative, the second-order tortuosity coefficient $w_{eff}$ >0 and $G' = -\mu_{eff} / (2w_{eff}) > 0$. Therefore, when the tip is pressed to apply a negative strain gradient, the hysteresis electric field $V_{d2}$ will first increase. Then decrease after the strain gradient exceeds $G'$. The corresponding absolute value of the total polarization intensity $P_2$ maintains the same trend. The extreme point $G'$ was obtained through fitting to be approximately $1.2\times10^6$ m$^{-1}$, which is comparable to the theoretical value.

## VII. Strain gradients and flexoelectricity for cantilever system

The strain gradient in a cantilever system can be written as:

$$\frac{\partial \varepsilon_{11}}{\partial x_3} = \frac{3w(L)}{L^2}\left(1 - \frac{x}{L}\right) \tag{M7.1}$$

where $w(L)$ is the vertical deflection defection delivered by the piezoelectric actuator at the end of the cantilever, which was collected by a laser displacement sensor. $L$ is the length of the cantilever and $x$ is the horizontal position along the cantilever.

When the change in curvature (strain gradient) with time during mechanical vibration can be expressed as $G=G_0\sin2\pi f_v t$, the flexoelectric polarization in the materials can be expressed as

$$P_f = w_{eff}(G_0 \sin 2\pi f_v t)^2 + \mu_{eff} G_0 \sin 2\pi f_v t \tag{M7.2}$$

The flexoelectric signal was measured through a dynamic thermomechanical analysis system (DMA850, TA Instruments, USA). We used DMA to deliver an oscillatory bending, The displacement signal is collected by a laser displacement sensor (KEYENCE, LK-G30, Japan). According to the natural plane assumption of the beam, the bending makes the material be stretched above the neutral plane, and compressed under the neutral plane. That results in a strain gradient uniformly distributed along the thickness of the cantilever beam. The bend-introduced charges generated of the sample through the charge amplifier (SA-PE01, Shiao Technology, Wuxi, China) and were collected by the oscilloscope (MDO3024, Tektronix, USA). Therefore, the polarization can be obtained by the total electric charge collected on the electrode:

$$P_3 = \frac{Q'}{S} \tag{M7.3}$$

where $Q'$ is the charge, $S$ is the electrode area.

## VIII. Polarization form through electrode topology design

For the capacitor poled parallel to the strain gradient (capacitor 1) the relationship between the bending-induced polarization and strain gradient is (Fig. S15a)

$$P_f(1) = wG^2 + \mu G \tag{M8.1}$$

Meanwhile, for the capacitor poled antiparallel to the strain gradient (capacitor 2), the relationship between the flexo-polarization and the strain gradient is (Fig. S15a)

$$P_f(2) = -wG^2 + \mu G \tag{M8.2}$$

If the a and c, b and d electrodes are joined (Abbreviation: ac-bd), the total flexoelectric polarization induced by bending is the sum (Fig. S15b)

$$P_f = \frac{P_f(1)+P_f(2)}{2} = \mu G \tag{M8.3}$$

Instead, if the a and d, b and c electrodes are joined, the total flexoelectric polarization induced by bending is the sum (Fig. S15b):

$$P_f = \frac{P_f(1)-P_f(2)}{2} = wG^2 \tag{M8.4}$$

Supplementary Materials for

# Second Harmonic Flexoelectricity

Zhiguo Wang [1,2,3,11], Yida Yang [4,11], Shuai Yang [5], Haoran Li [1], Chunchun Li [5], Zhibin Wen [1,2,3], Zhengqiu Xie [6], Wenbin Huang [7], Konstantin Shapovalov [8], Jiawang Hong [4],

Longlong Shu [1,2,3]✉, Fei Li [5]✉, Gustau Catalan [9,10]✉

[11.] *School of Physics and Materials Science, Nanchang University, Nanchang 330031, Peoples R China.*

[12.] *Jiangxi Provincial Center for Interdisciplinary Studies in Future Display, Nanchang University, Nanchang 330031, Peoples R China.*

[13.] *The Jiangxi Provincial Engineering Technology Research Center for Advanced Functional Film Materials, Nanchang University, Nanchang 330031, Peoples R China.*

[14.] *Beijing Inst Technol, Sch Aerosp Engn, Beijing 100081, Peoples R China*

[15.] *Xi An Jiao Tong Univ, State Key Lab Mech Behav Mat, Educ Minist, Elect Mat Res Lab,Key Lab, Xian, Shaanxi, Peoples R China*

[16.] *College of Mechanical Engineering, Chongqing University of Technology, Chongqing 400050, Peoples R China*

[17.] *Chongqing Univ, State Key Lab Mech Transmiss, Chongqing 400044, Peoples R China*

[18.] *Univ Liege, Theoret Mat Phys, Q MAT, Sart Tilman Par Liege, Belgium*

[19.] *Institucio Catalana de Recerca i Estudis Avançats (ICREA), Barcelona, Catalonia.*

[20.] *Institut Catala de Nanociencia i Nanotecnologia (ICN2), CSIC–BIST, Campus Universitat Autonoma de Barcelona, Barcelona, Catalonia.*

[11.] *These authors contributed equally*: Zhiguo Wang, Yida Yang.

✉*e-mail*: llshu@ncu.edu.cn; ful5@xjtu.edu.cn; gustau.catalan@icn2.cat

## S1. SEM and XRD for $BaTiO_3$

The SEM images were recorded on the polished and thermally etched surfaces of $BaTiO_3$ ceramic (Fig. S1-1). The sample exhibited dense microstructures with minimal visible porosity. As quantified using Nano Measurer software, the average grain size of $BaTiO_3$ was approximately 12 μm.

The sample is sintered densely (Method II). The Rietveld refinements were performed to confirm the phase purity and examine the detailed evolution of the lattice parameters. An initial model for the tetragonal $ABO_3$-type perovskite with a P4mm space group was adopted. The measured XRD data exhibit excellent agreement with the calculated results, thereby confirming the reliability of the structural simulation.

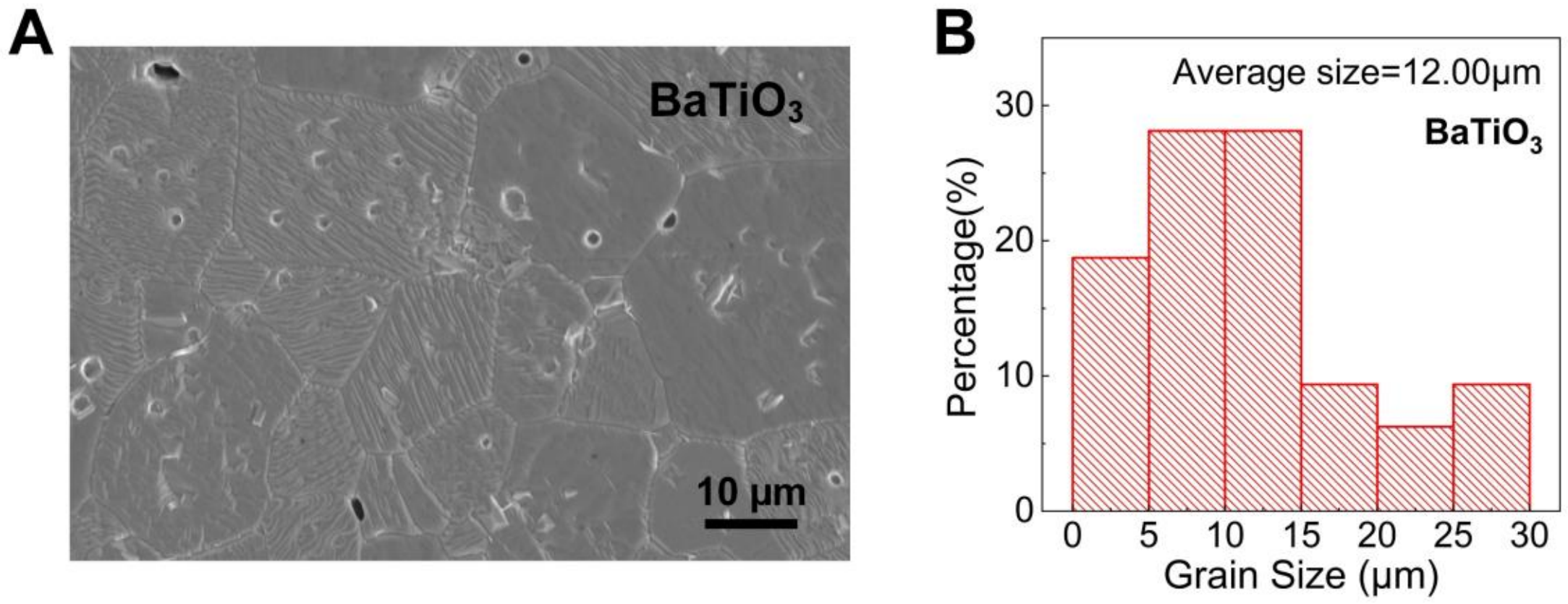

**Figure S1-1**. (**A**) SEM for $BaTiO_3$. (**B**) Grain size for $BaTiO_3$.

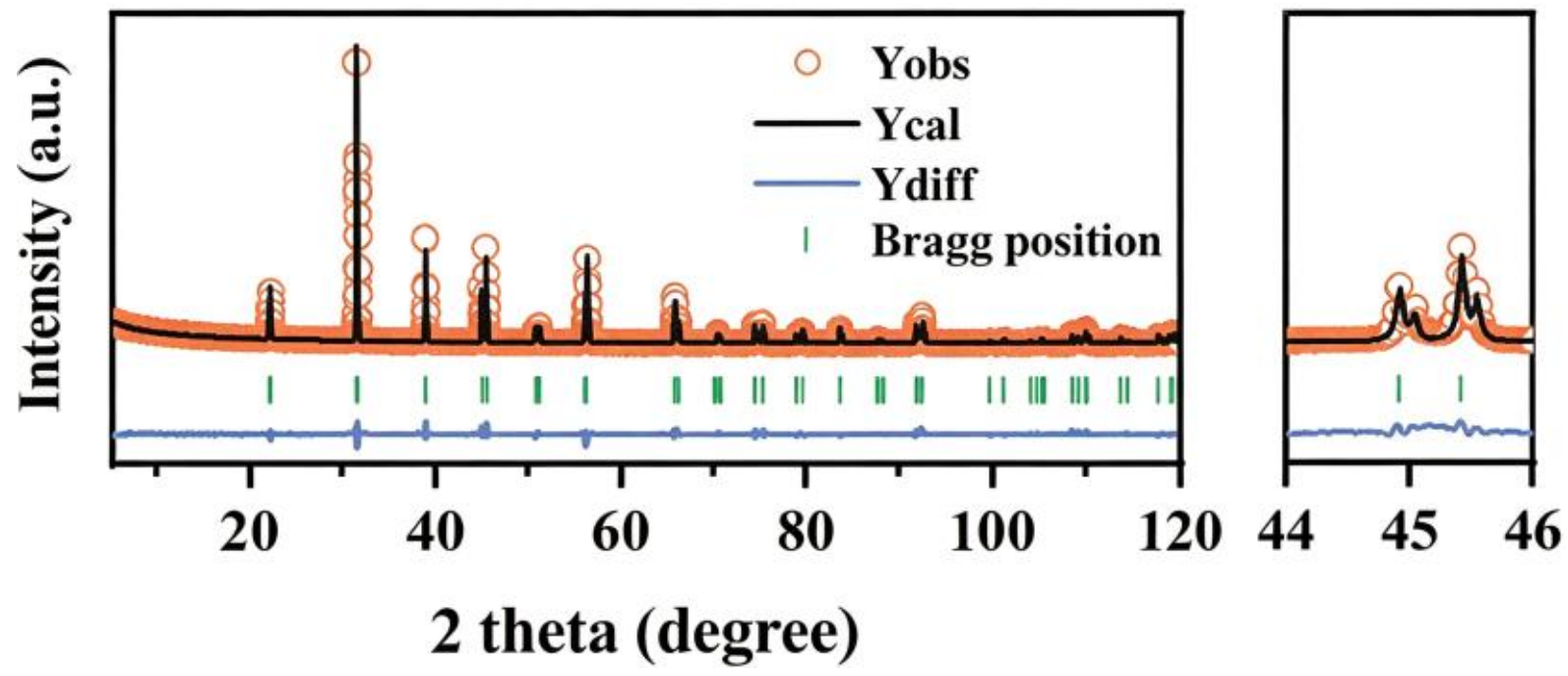

**Figure S1-2**. XRD for $BaTiO_3$.

## S2. The parameters of $BaTiO_3$

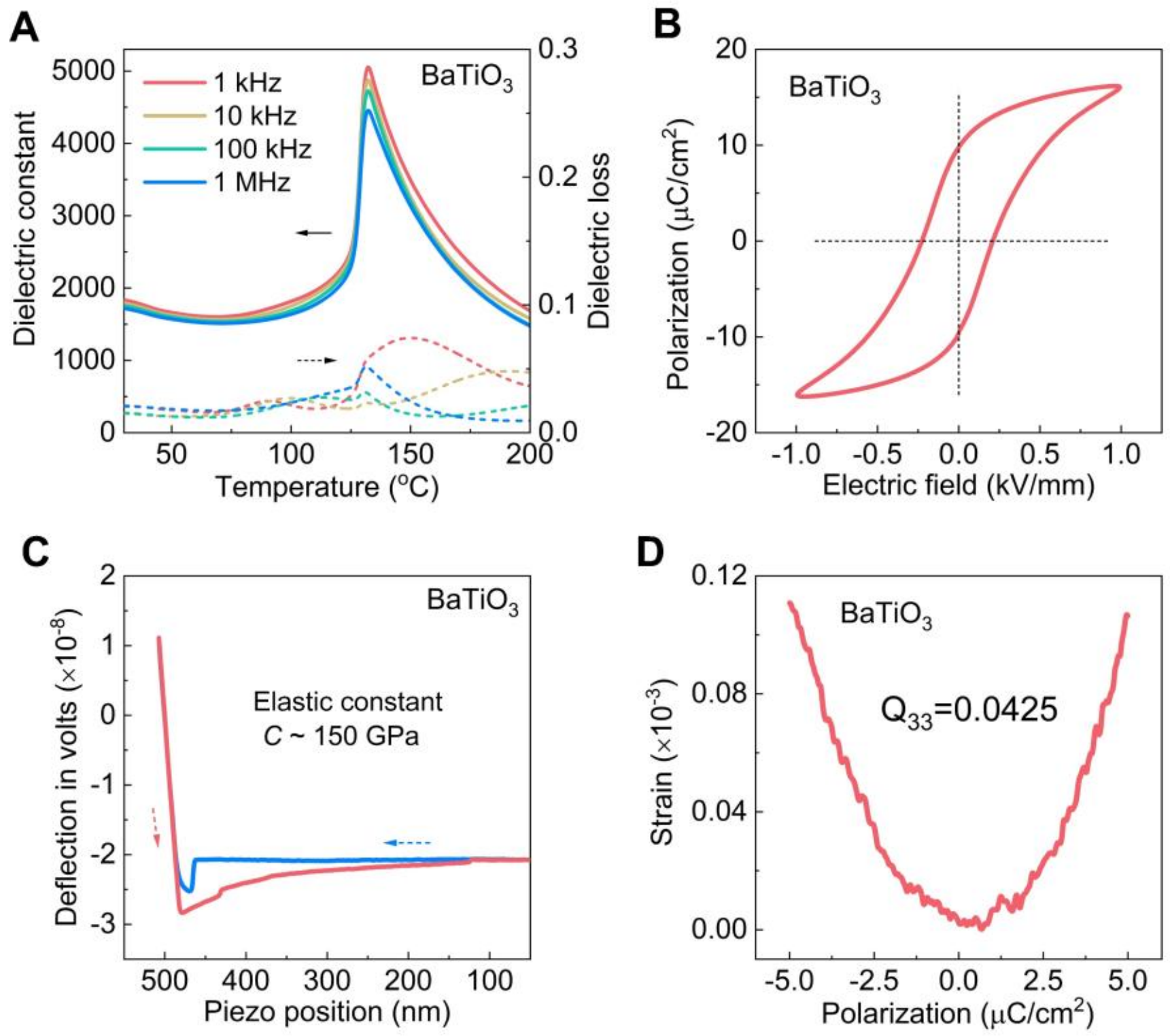

**Figure S2.** Material parameters of $BaTiO_3$ used to calculate the second-order flexoelectric coefficient: (**A**) temperature-dependent dielectric constant, (**B**) polarization-electric field hysteresis loop yielding $P_0$, (**C**) elastic constant $C$, and (**D**) electrostriction coefficient $Q_{33}$. The longitudinal electrostrictive coefficient $Q_{zz}$ is related to the transverse one by $Q_{31}/Q_{33} = \nu$ . Since the electrostrictive coefficient is directly proportional to the piezoelectric coefficient (see Methods I), so $\nu = d_{31}/d_{33}$, where ν is effective piezoelectric Poisson's ratio. Such as, for poled BTO *ceramic*, ν ~ 0.6 - 0.8. For BTO *single crystal*, ν ~ 0.3 - 0.4 [S1].

## S3. The thickness of $BaTiO_3$ film

Since the BTO film deposited with 300 pulses is too thin to allow accurate thickness fitting directly from X-ray reflectivity (XRR), a linear relationship between the pulse number and film thickness was established using thicker samples. The linear fitting results show that 1000 pulses correspond to a film thickness of approximately 29 nm. Based on this relationship, the thickness of the BTO film deposited with 300 pulses is estimated to be approximately 8.5 nm.

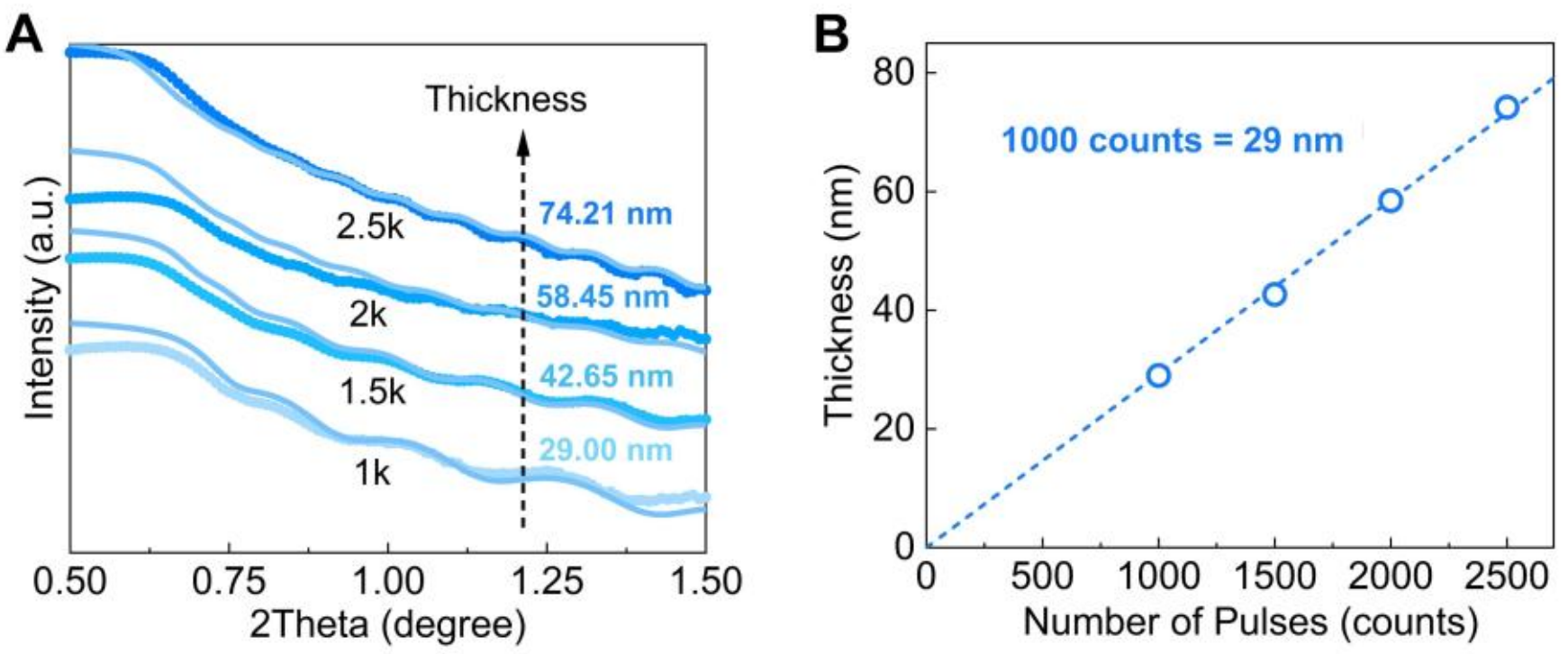

**Figure S3**. Thickness calibration of BTO films prepared with different numbers of laser pulses. (**A**) XRR profiles and fitting curves of BTO films prepared with different pulse numbers; (**B**) Dependence of the BTO film thickness on the number of laser pulses.

### S4. Structural characterization for BTO/SRO/STO/mica

Fig. S4A and S4B show a typical out-of-plane $2\theta/\omega$ scan of the BTO/SRO/STO/mica heterostructure. On the mica(001) substrate, only the BTO and SRO (111) and (222) diffraction peaks are observed, indicating that the BTO and SRO films exhibit a highly preferred out-of-plane orientation, with no obvious secondary-phase peaks detected. The BTO(111) peak of the film grown on the mica substrate is close to the bulk BTO value, suggesting that the film is in a relaxed strain state. This conclusion is further supported by the reciprocal space mapping (RSM) results. Fig. S4C shows the reciprocal space map of the heterostructure around the asymmetric BTO(121) and SRO(121) reflections. The BTO(121) and SRO(121) diffraction peaks are nearly aligned along the in-plane $Q_x$ direction, indicating a good in-plane coherent relationship between the BTO film and the SRO bottom electrode layer.

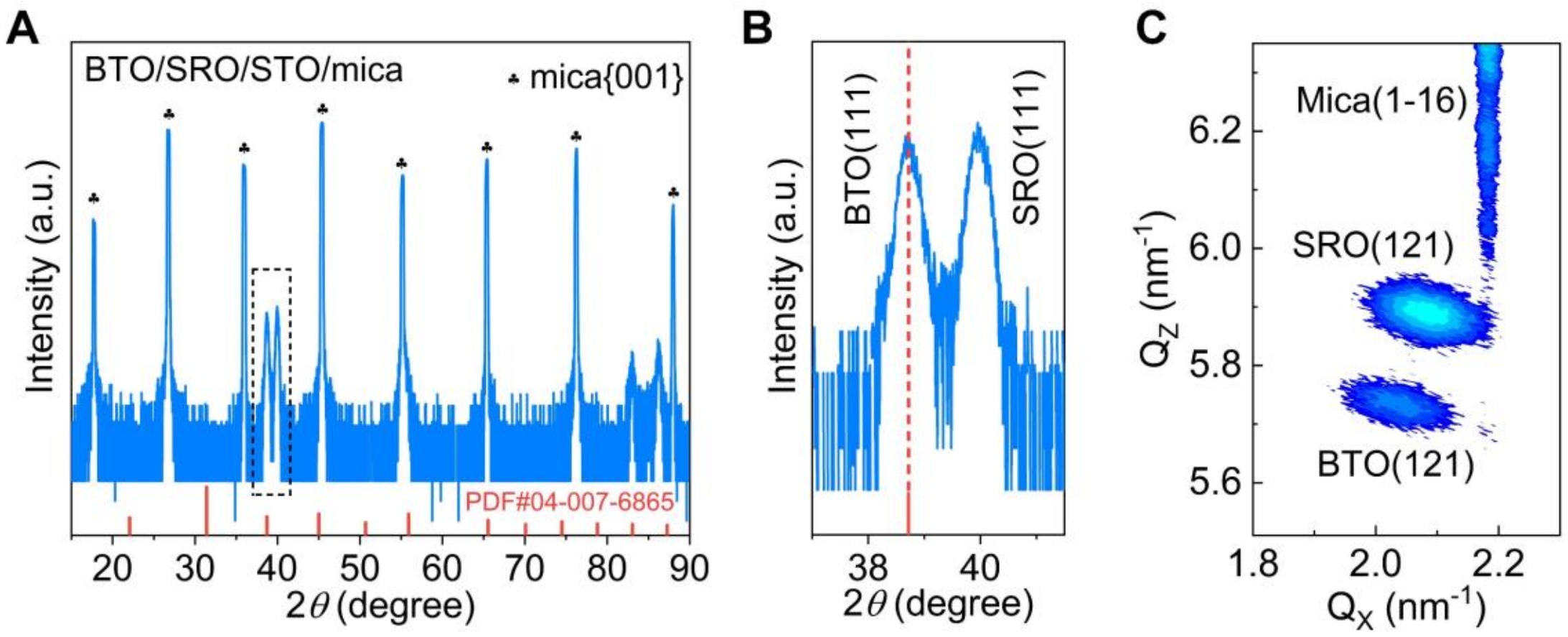

**Figure S4**. Structural characterization of the BTO/SRO/STO/mica heterostructure. (**A**) Typical out-of-plane $2\theta/\omega$ scan; (**B**) Enlarged view of the BTO(111) and SRO(111) diffraction peaks; (**C**) Reciprocal space map around the asymmetric BTO(121) and SRO(121) reflections.

## S5. Ferroelectricity for $BaTiO_3$ film with mica substrate

The ferroelectric switching behavior of BTO thin films grown on STO and mica substrates was investigated by Piezoelectric Force Microscopy (PFM). Fig. S5 in supplementary show the PFM phase and amplitude images after domain writing. A box-in-box domain pattern was written by applying different DC biases to the conductive AFM tip. First, a 3×3 $\mu m^2$ region was poled by scanning with a +6 V tip bias. Subsequently, the central 2×2 $\mu m^2$ region was switched by applying -6 V, followed by writing the central 1×1 $\mu m^2$ region again with +6 V. The PFM phase image exhibits a clear 180° contrast between oppositely poled regions, while the corresponding amplitude image shows distinct contrast at the domain boundaries, confirming the reversible ferroelectric polarization switching of the BTO thin film.

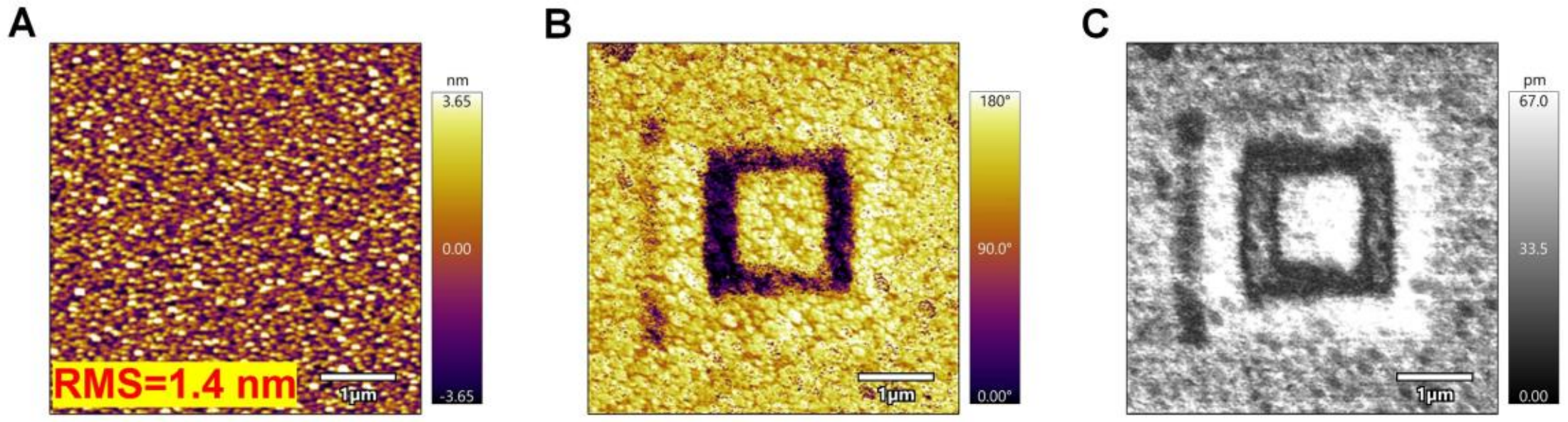

**Figure S5**. (**A**) AFM topography images of the BTO/SRO/STO/mica heterostructures; PFM out-of-plane amplitude and phase images electrical poling. (**B**) Phase, (**C**) Amplitude images of BTO/SRO/STO/mica heterostructure.

## S6. The relationship for AFM tip force and strain gradient of $BaTiO_3$ film

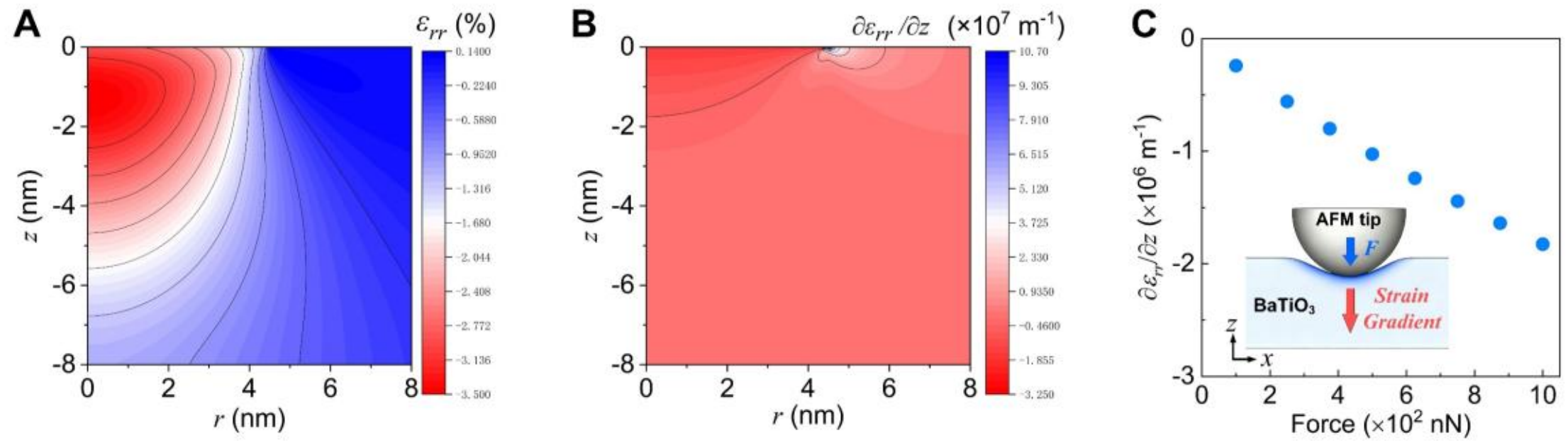

**Figure S6**. Through COMSOL, a force simulation analysis was conducted on the BTO film, and the relationship between the strain gradient and the force applied by the AFM tip was obtained.

## S7. PFM measurements for $BaTiO_3$ film

Subsequently, switching spectroscopy piezoelectric force microscopy (SS-PFM) measurements were performed on the BTO thin film. A fixed position on the film surface was selected, and different normal loads ranging from 100 nN to 1000 nN were applied to the AFM tip. Under the same dc bias condition, PFM phase hysteresis loops were recorded at each applied load to investigate the influence of mechanical force on the ferroelectric switching behavior (Fig. S7).

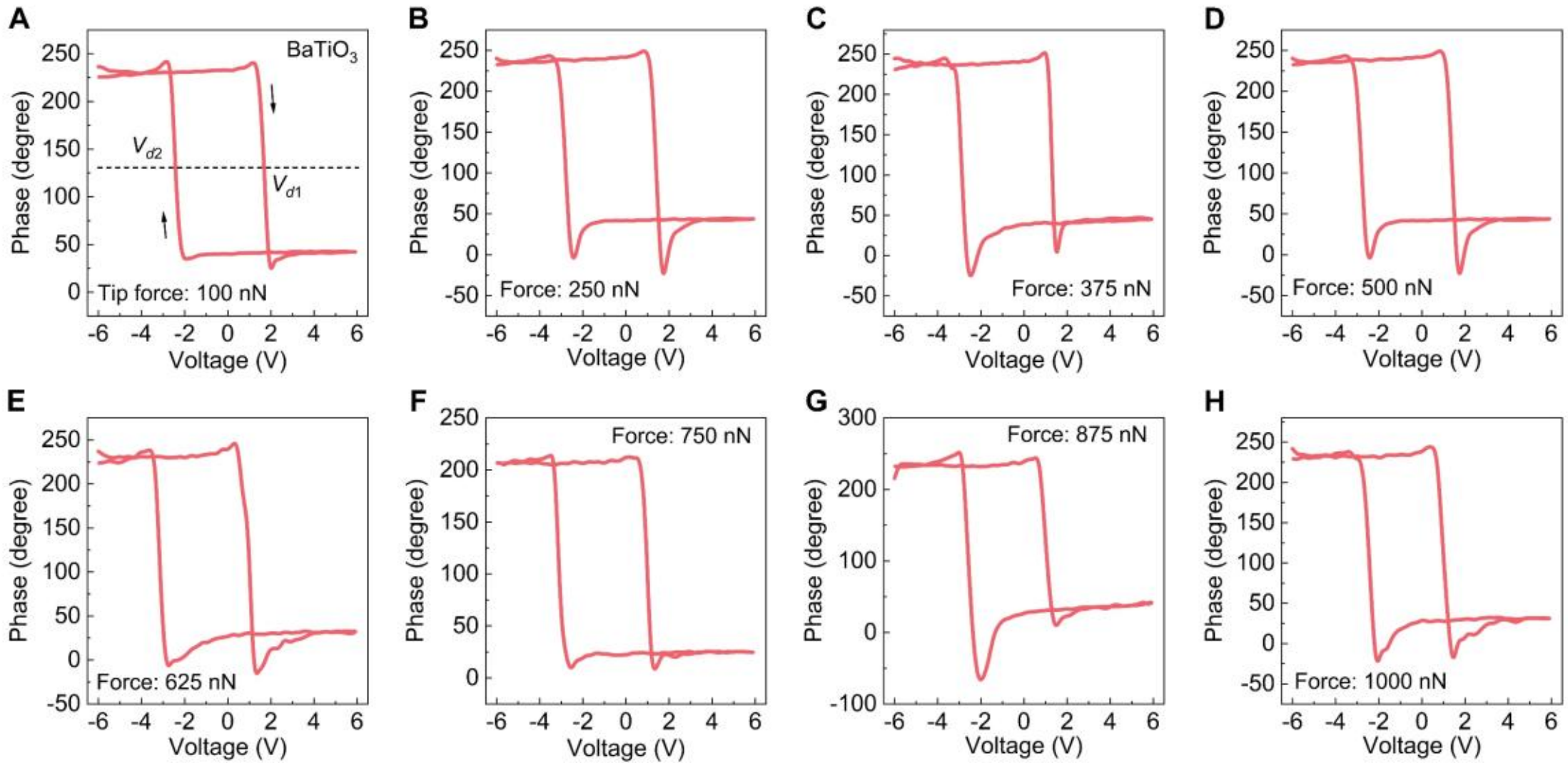

**Figure S7**. PFM phase hysteresis loops were recorded at applied load from 100 nN to 1000 nN.

## S8. Flexoelectric coefficient measurement device

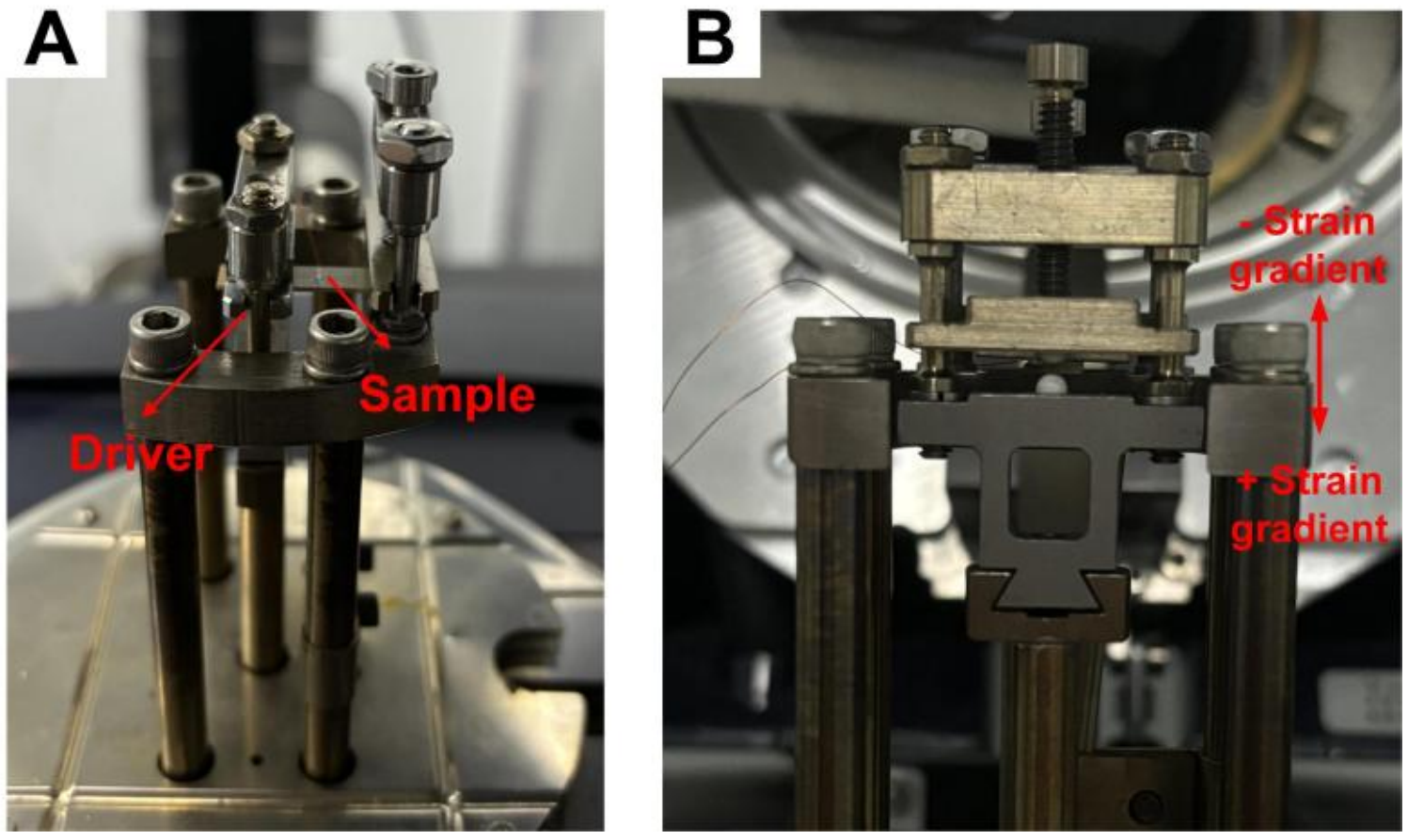

**Figure S8.** (**A**) The end of the sample is clamped, and the front end is driven by an actuator to undergo reciprocal cyclic motion. (**B**) When the sample bends downwards, the top surface is subjected to tension while the bottom surface is compressed, forming a positive strain gradient. Conversely, when the sample bends upwards, a negative strain gradient is formed.

The flexoelectric coefficients were measured through a dynamic thermomechanical analysis system (DMA850, TA Instruments, USA). We used DMA to deliver an oscillatory bending. According to the natural plane assumption of the beam, the bending makes the material be stretched above the neutral plane, and compressed under the neutral plane. That results in a strain gradient uniformly distributed along the thickness of the cantilever beam.

## S9. As the strain gradient increases, the waveform changes in real time for $BaTiO_3$

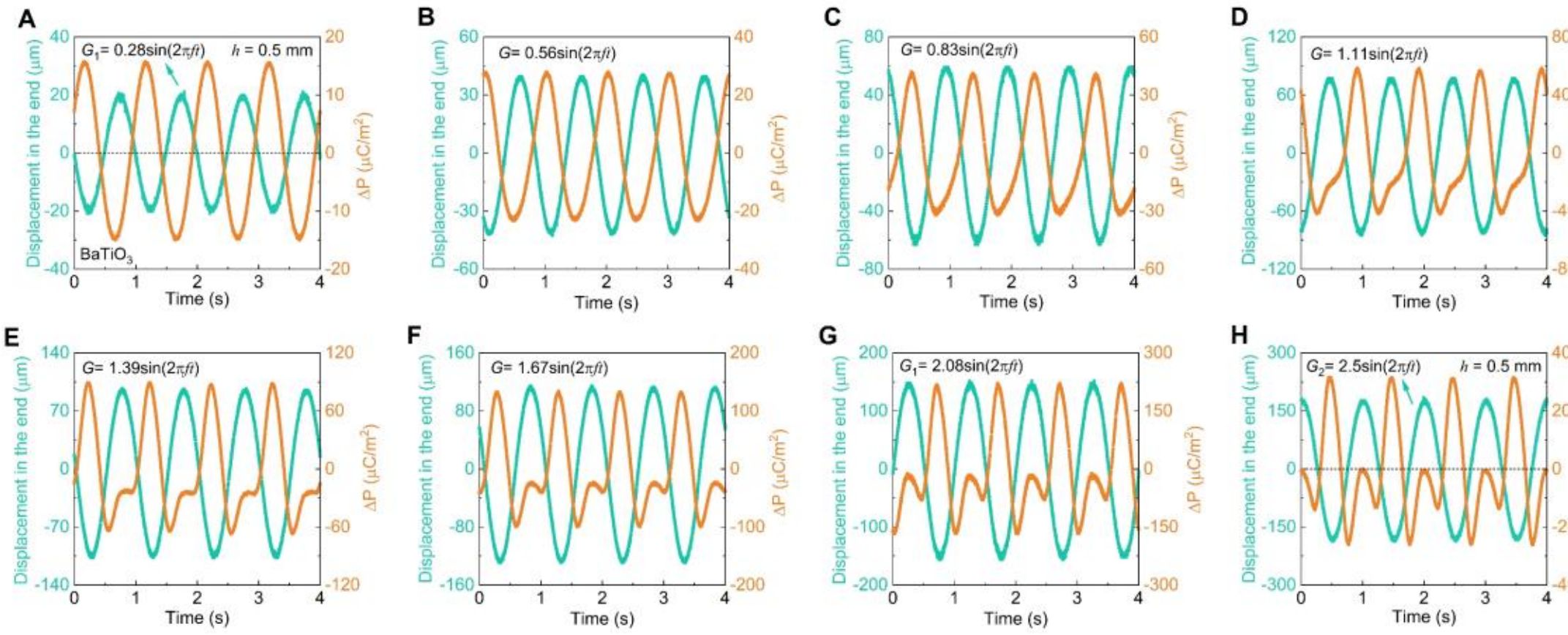

**Figure S9-1** For BTO ceramic with thicknesses of 0.5 mm. The strain gradient gradually increased from 0.28 $m^{-1}$ to 2.5 $m^{-1}$, the second-order information of the output signal gradually intensified.

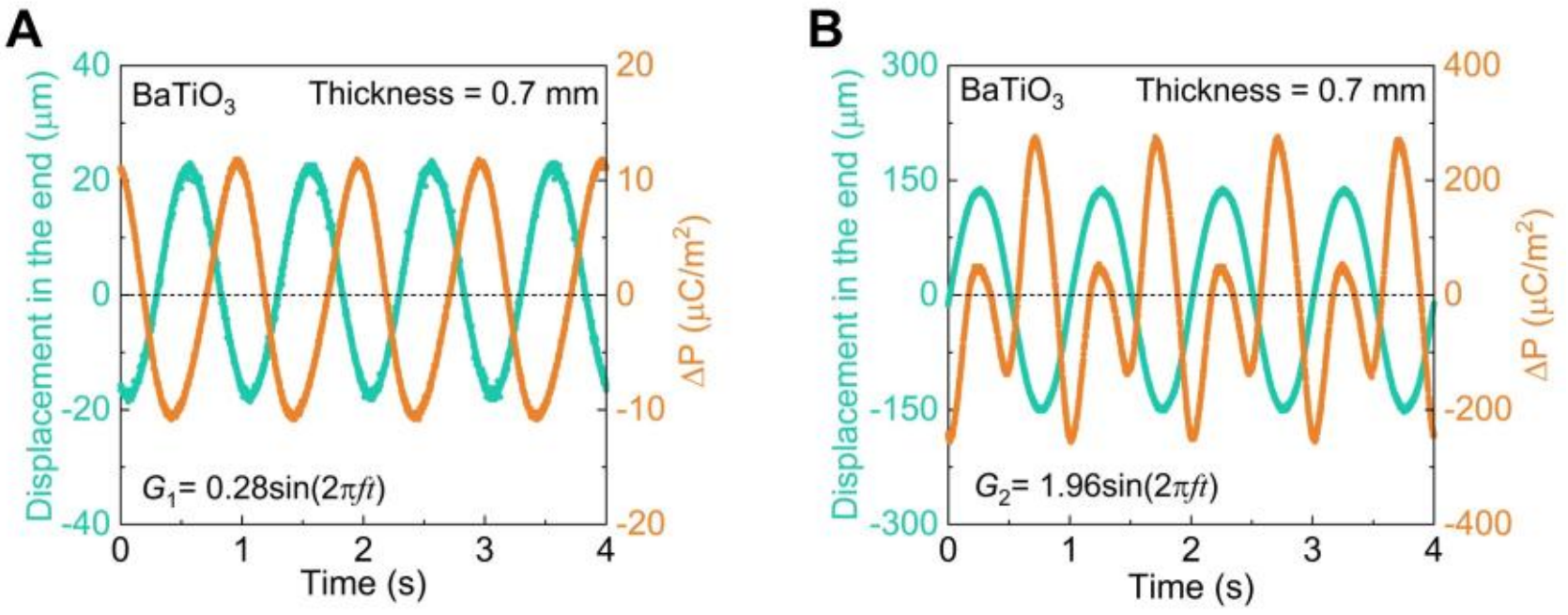

**Figure S9-2.** For BTO ceramic with thicknesses of 0.7 mm. (**A**) in a very small strain gradient range ($G_1 < G'$), the output electrical signal is at the same frequency as the vibration. (**B**) The frequency-doubling response observed under large strain gradients is demonstrated.

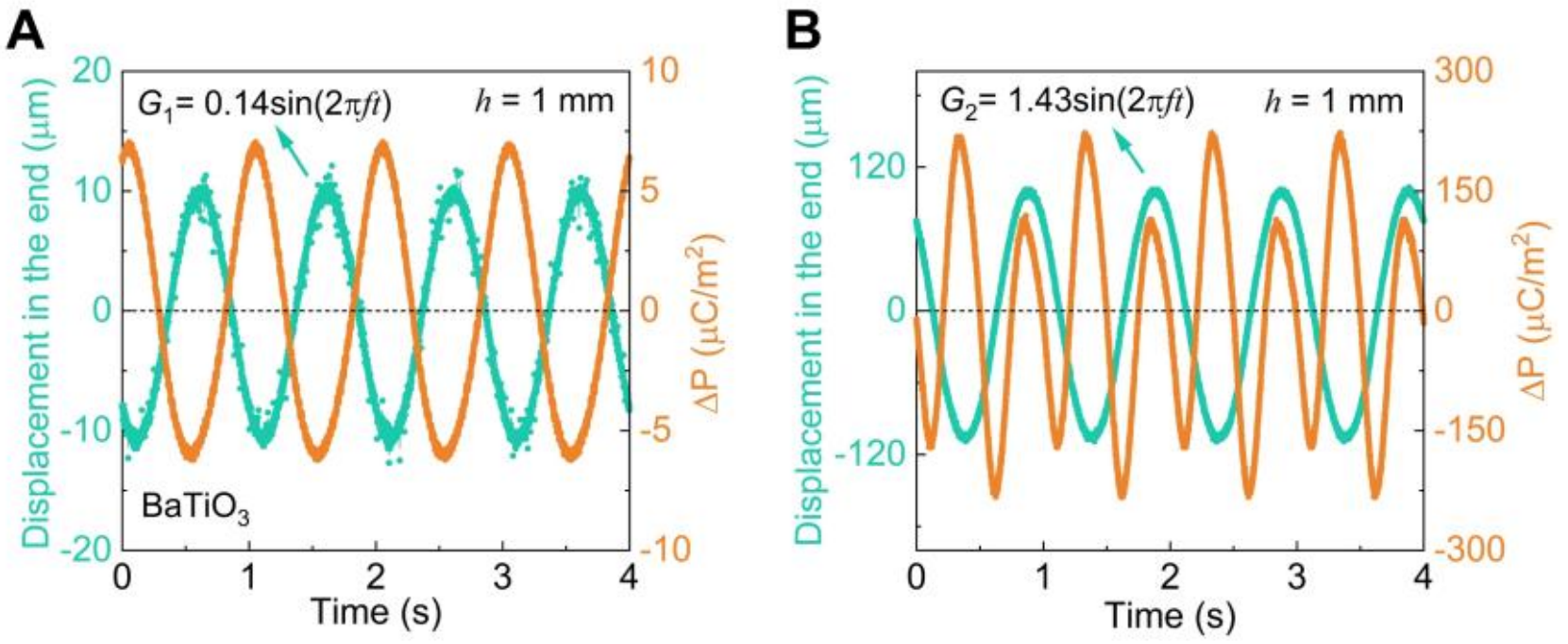

**Figure S9-3.** For BTO ceramic with thicknesses of 1 mm. (**A**) in a very small strain gradient range ($G_1 < G'$), the output electrical signal is at the same frequency as the vibration. (**B**) The frequency-doubling response observed under large strain gradients is demonstrated.

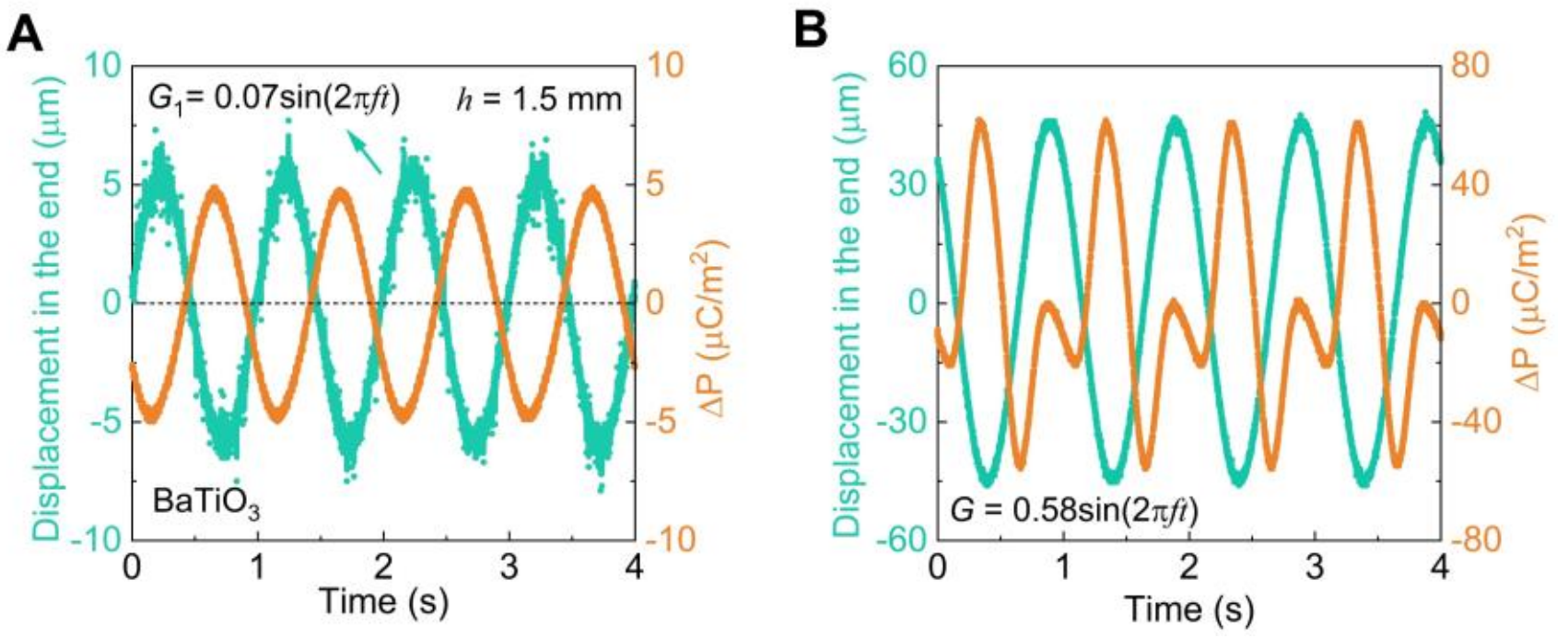

**Figure S9-4.** For BTO ceramic with thicknesses of 1.5 mm. (**A**) in a very small strain gradient range ($G_1 < G'$), the output electrical signal is at the same frequency as the vibration. (**B**) The frequency-doubling response observed under large strain gradients is demonstrated.

## S10. Fourier analysis for polarization intensity of BTO

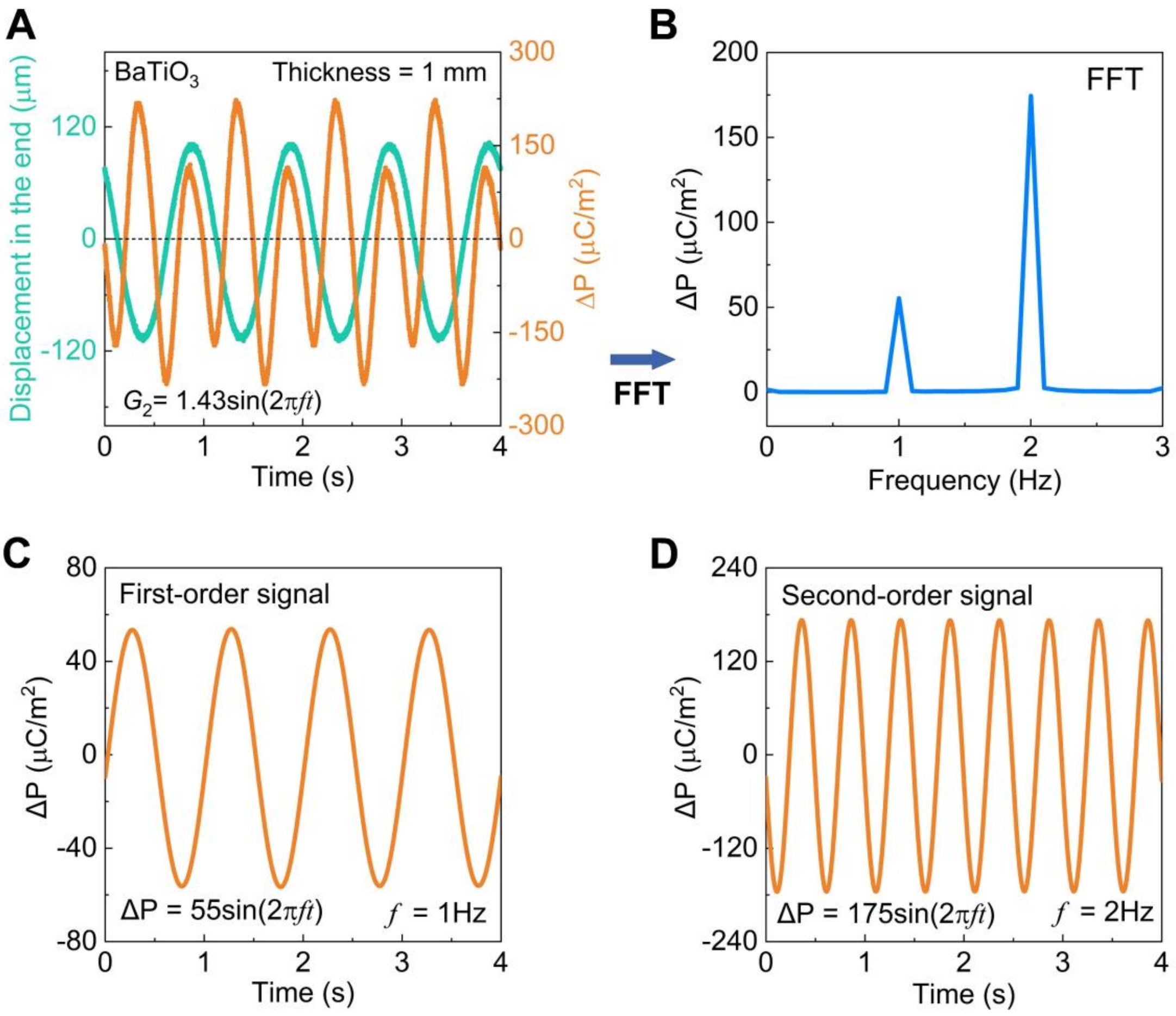

**Figure S10.** For BTO ceramic with thicknesses of 1 mm. (**A**) The frequency-doubling response observed under large strain gradients is demonstrated. (**B**) The By performing *Fourier* analysis, it can be determined that the curve is composed of frequency signals at 1 Hz and 2 Hz for polarization curve in (Fig. **A**). When the input vibrational signal is at 1 Hz, the polarization curve can be decomposed into (**C**) a first-order signal at 1 Hz, and (**D**) a second-order signal at 2 Hz.

## S11. The relationship between the second-order polarization and the strain gradient for different thicknesses of BTO.

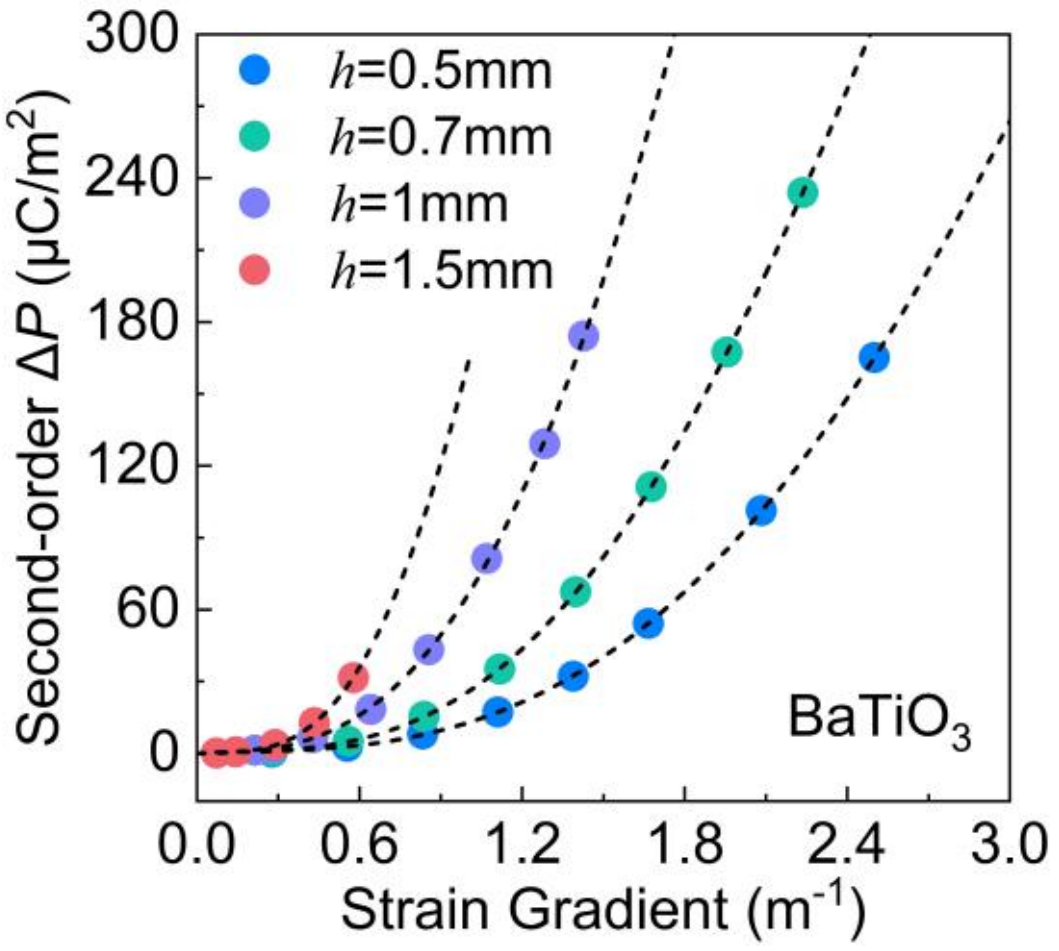

**Figure S11.** The relationship between the second-order polarization and the strain gradient for different thicknesses of BTO.

## S12. Temperature spectrum of output waveforms for $BaTiO_3$

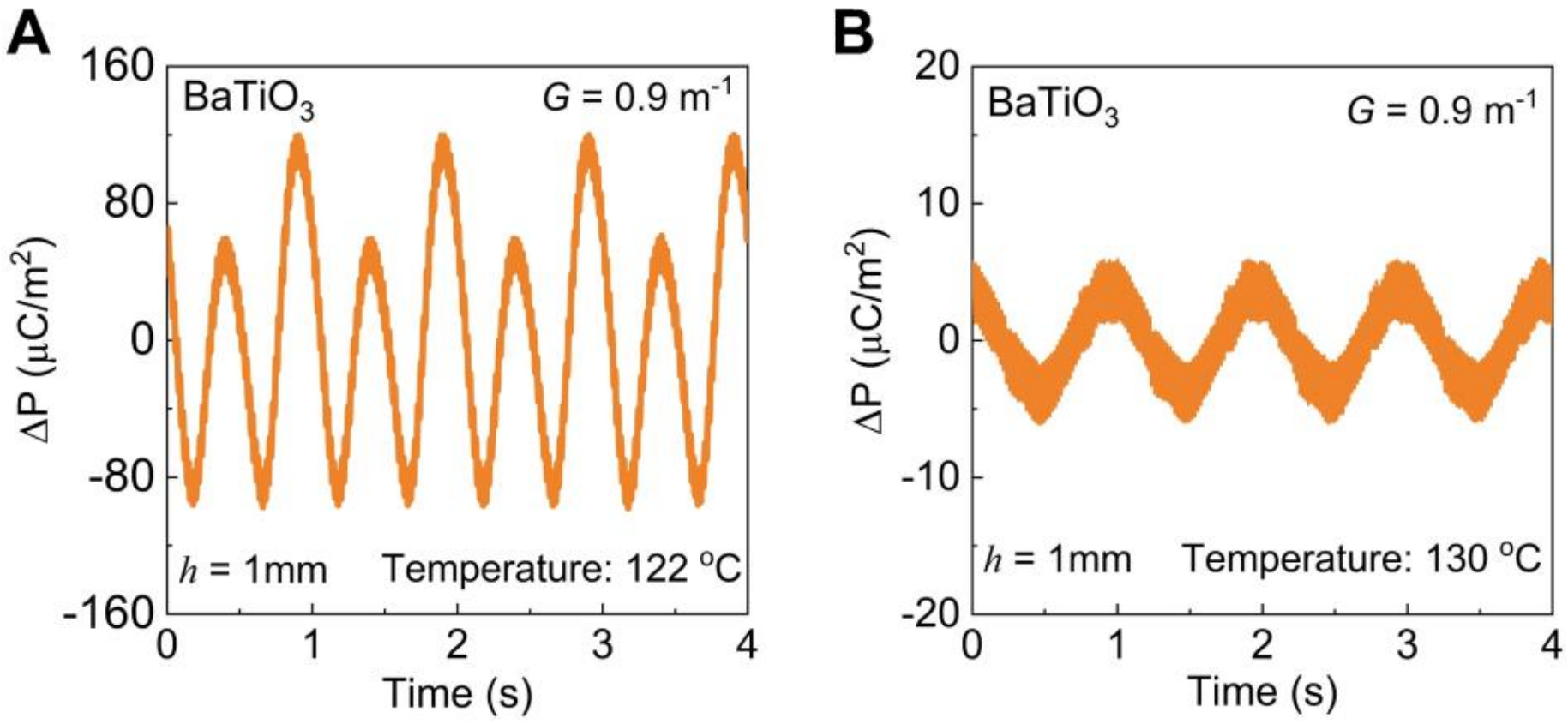

**Figure S12.** The output waveforms of BTO at (**a**) 122 ºC and (**b**) 130 ºC.

## S13. Nonlinear flexoelectricity in *x*Sm-PMN-PT

In order to extend the generality of the effect, another ferroelectric material, PMN-PT was selected, its piezoelectric coefficient was further increased by doping *x* mol% $Sm^{2+}$ ions (*x*Sm-PMN-PT, *x* = 0, 1, 2) [S2]. Figs. S13-1 to S13-3 presents the dielectric, ferroelectric properties and electrostrictive coefficients of *x*Sm-PMN-PT, respectively. The results (Fig. S13-4) show that, when the strain gradient has not passed the *G*', the polarization introduced by bending and the displacement are of the same frequency. Beyond *G*', the frequency of the bending-induced polarization is twice that of displacement. Similarly, with the increase of the initial piezoelectric coefficient, the strain gradient corresponding to the extreme value of the polarization curve becomes smaller. This result is the same as that of BTO-based materials. More importantly, for the 2Sm-PMN-PT material, the second-order coefficient reaches up to 10 times that of the first-order coefficient.

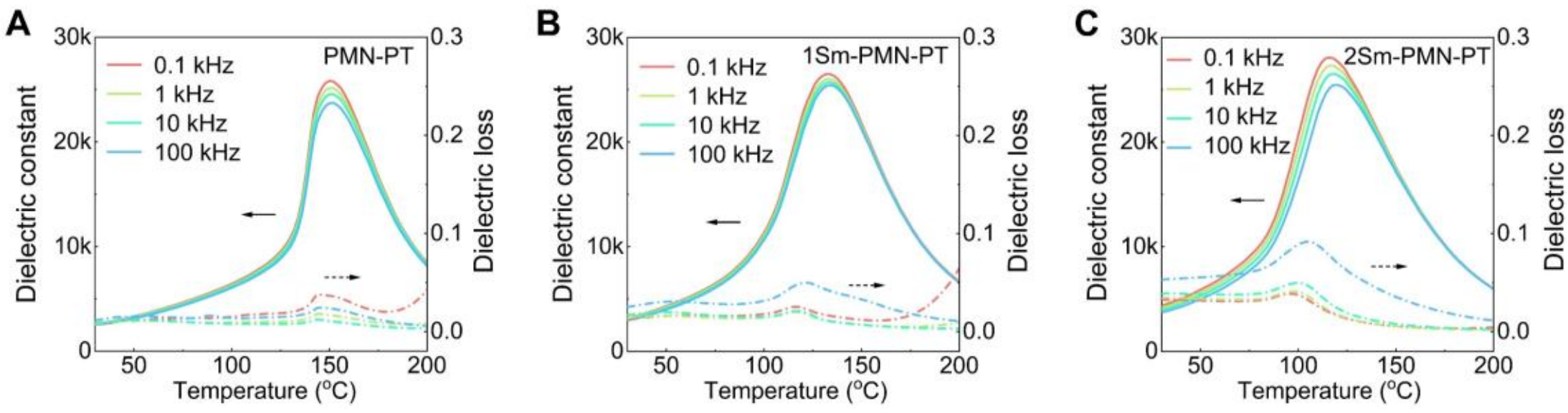

**Figure S13-1.** Temperature dependence of the dielectric constant and dielectric loss in the (**A**) PMN-PT, (**B**) 1Sm-PMN-PT and (**C**) 2Sm-PMN-PT ceramics.

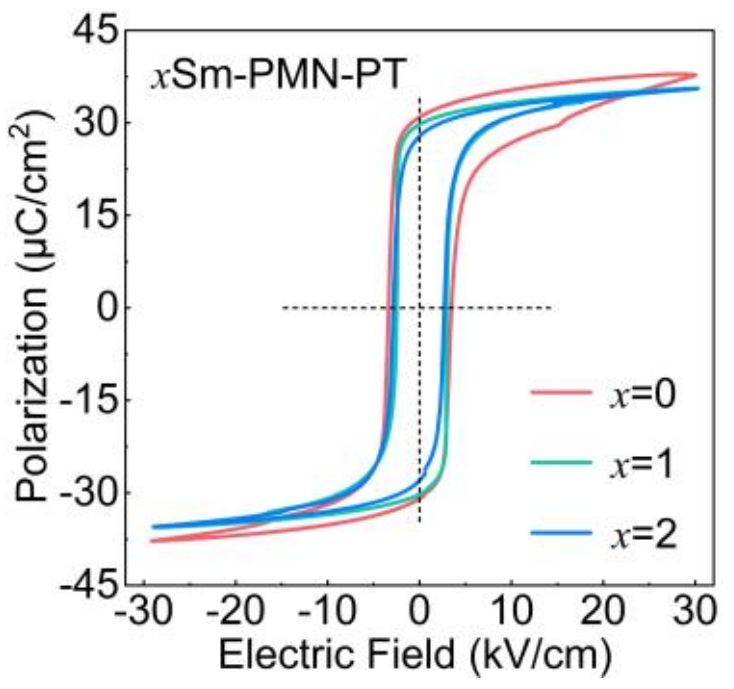

**Figure S13-2.** The P-E loop for *x*Sm-PMN-PT (*x* = 0, 1, 2).

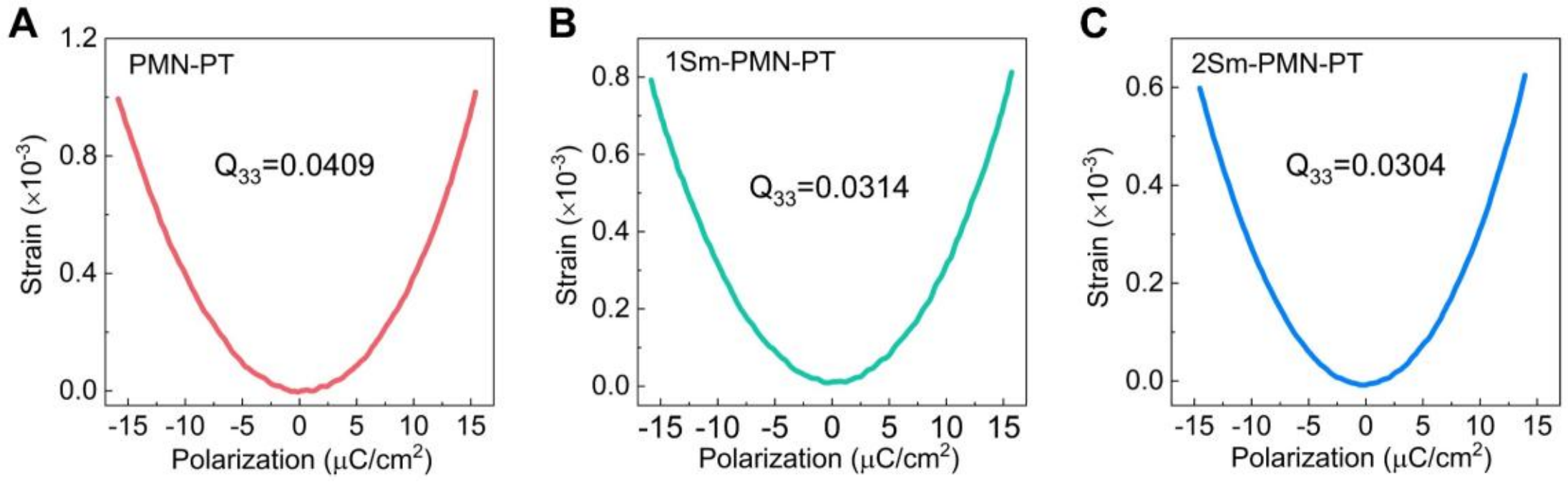

**Figure S13-3.** Electrostrictive performances for (**A**) pure PMN-PT, (**B**) 1Sm-PMN-PT, (**C**) 2Sm-PMN-PT.

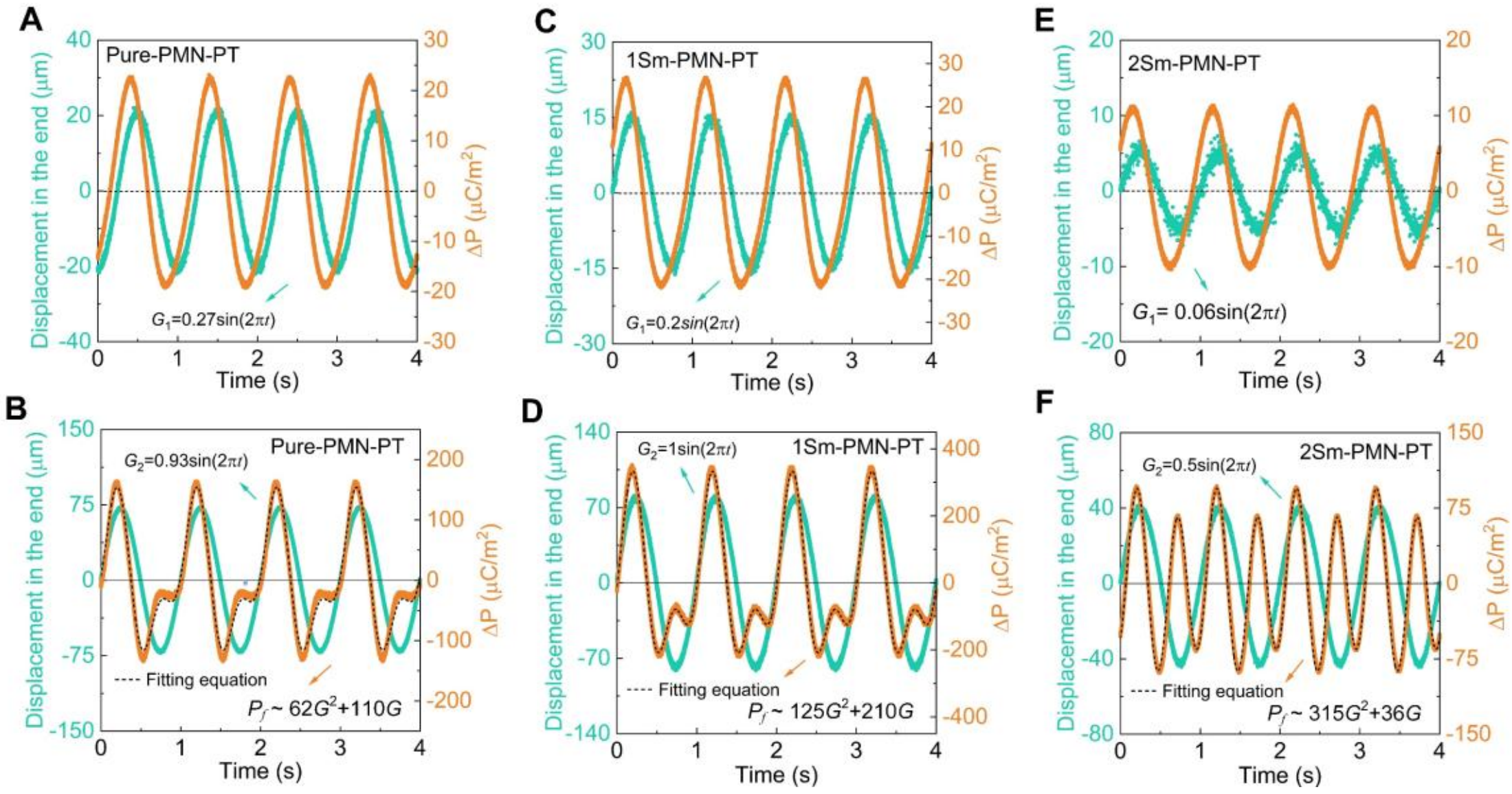

**Figure S13-4.** The real-time signals of displacement for (**A**, **B**) Pure PMN-PT, (**C**, **D**) 1Sm-PMN-PT and (**E**, **F**) 2Sm-PMN-PT.

## S14. Nonlinear flexoelectricity in PMN-PT single crystals

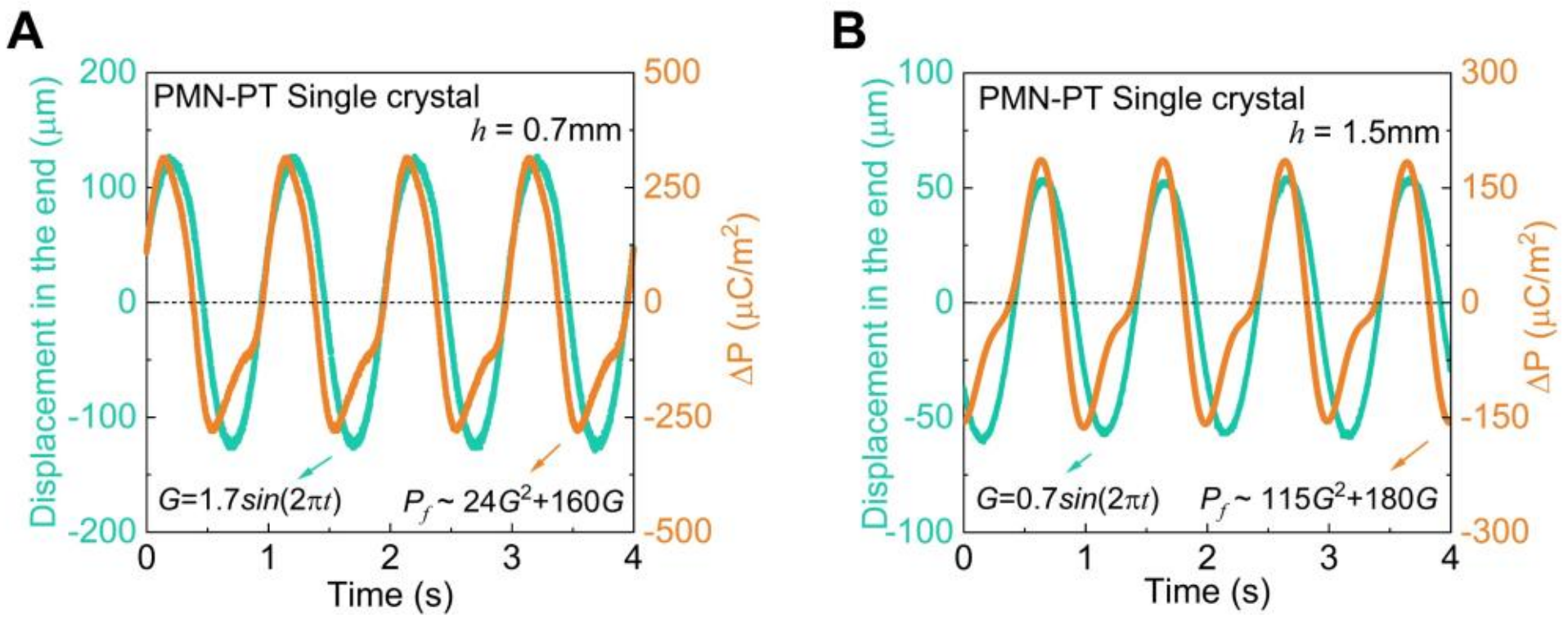

**Figure S14.** The real-time signals of displacement for PMN-PT single crystal with thickness of (**A**) 0.7mm and (**B**) 1.5mm. The second-order coefficient increases as the thickness increases.

## S15. The relationship between the flexoelectric polarization and the strain gradient for the topological electrode structures

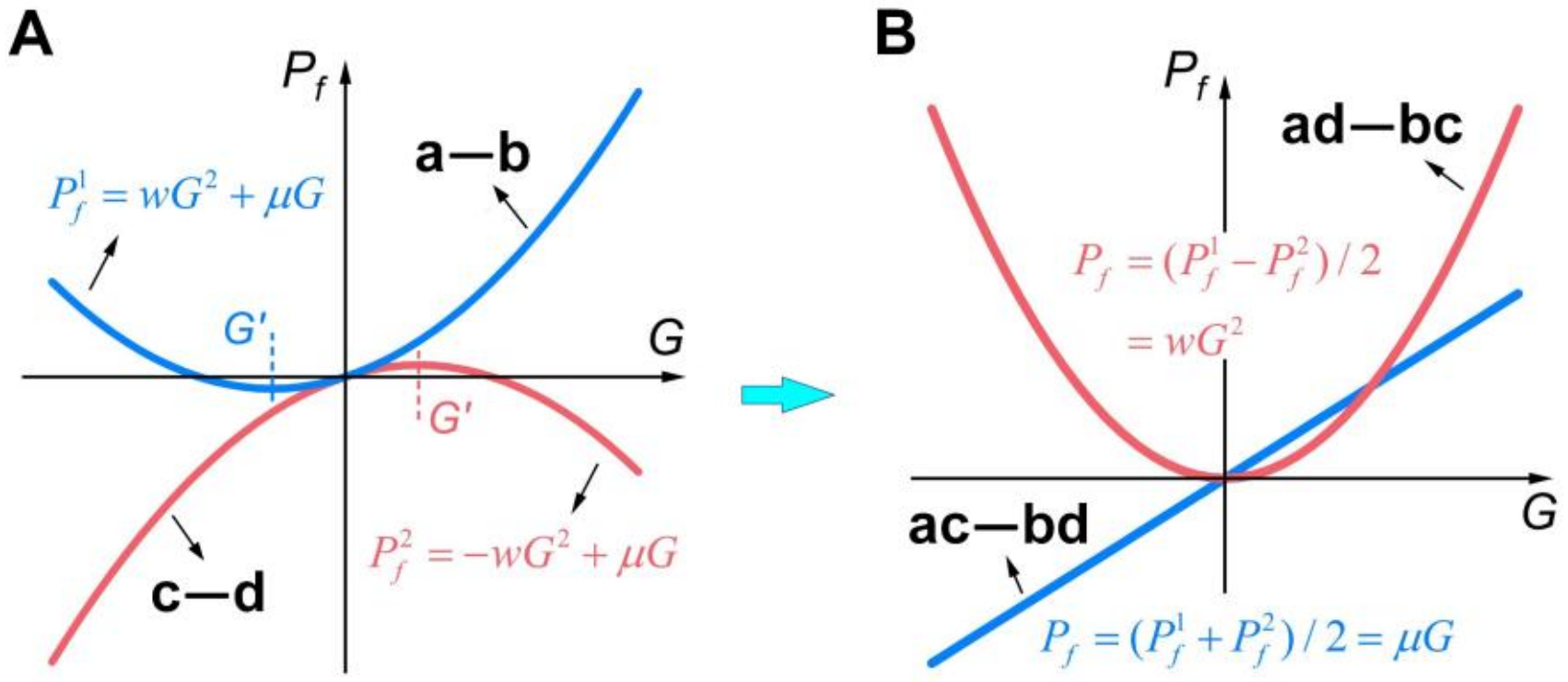

**Figure S15.** (**A**) Due to the inverse symmetry of the structure, when the initial prepolarization direction is opposite, the flexo-polarization induced by bending has opposite higher order term coefficients, while the linear term coefficients are the same. (**B**) When the ac-bd and ad-bc electrodes are connected in series, the linear effects and high-order of flexo-polarization on the strain gradient are realized respectively.

## S16. Real-time Signals for $BaTiO_3$ with topology design

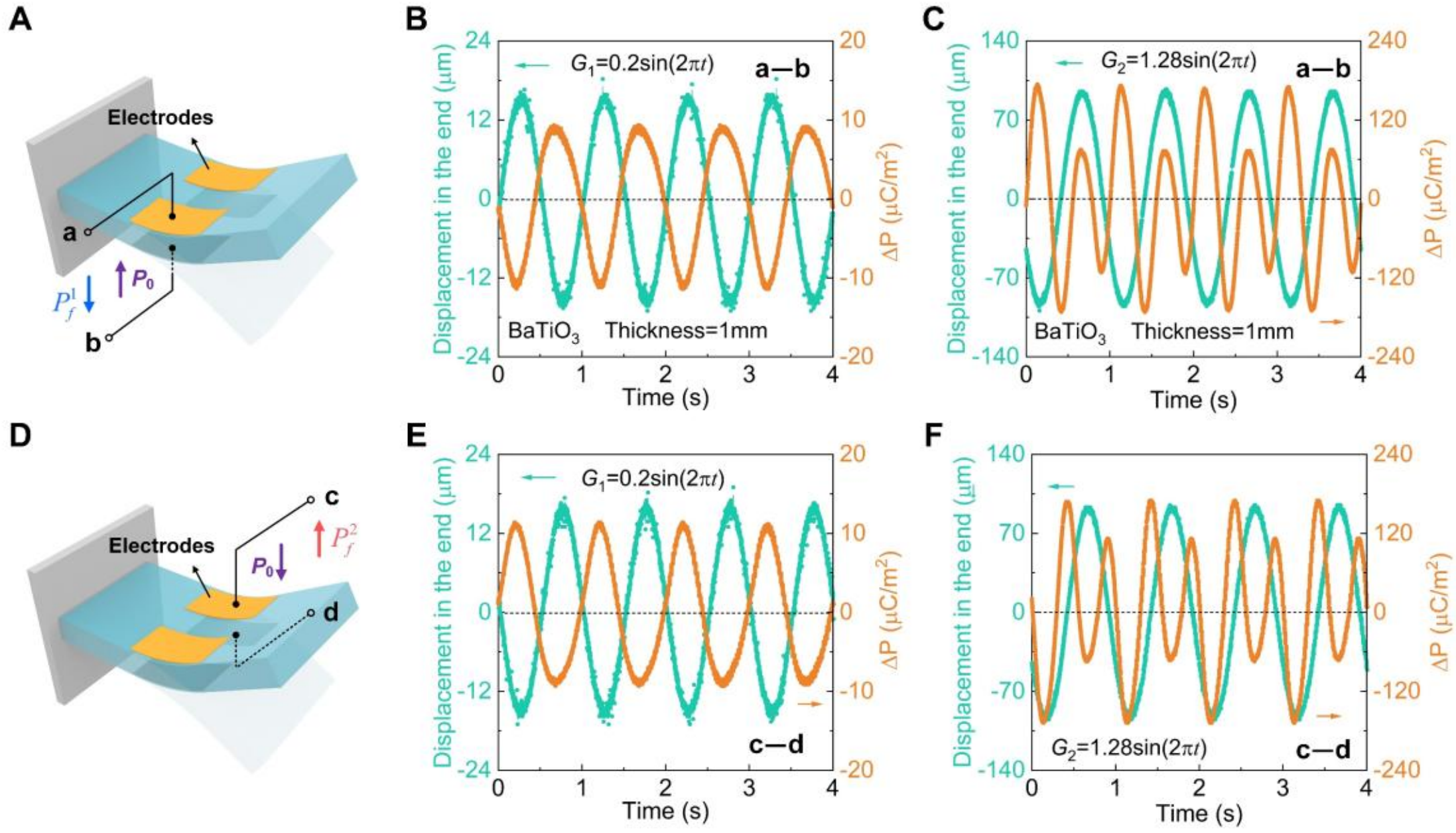

**Figure S16.** For $BaTiO_3$ with topological electrode structures, the real-time output of electrical signals and vibration displacements when connecting a single-sided electrode.

## S17. Real-time Signals for 2Sm-PMN-PT with topology design

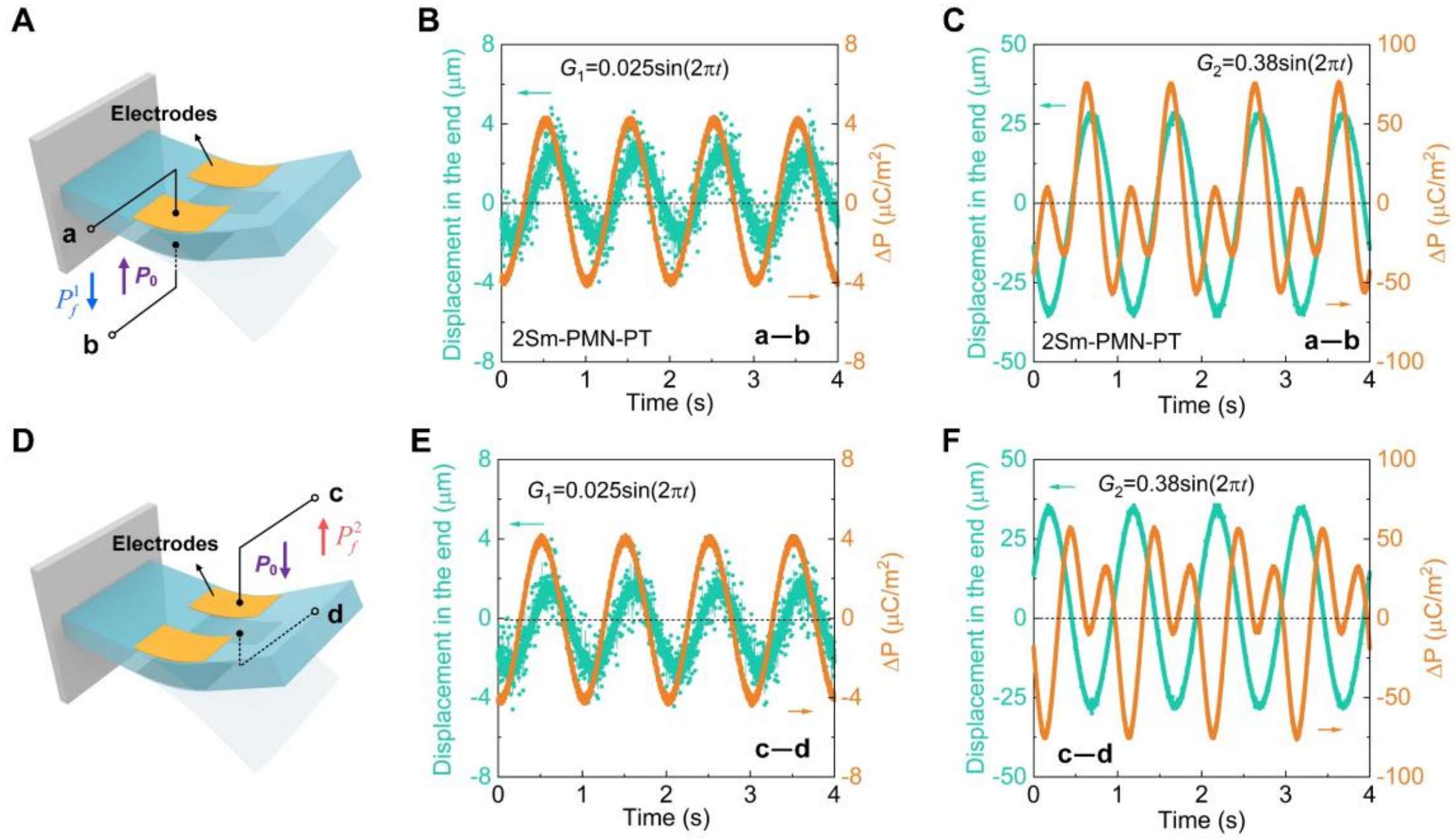

**Figure S17-1** For PMN-PT with topological electrode structures, the real-time output of electrical signals and vibration displacements when connecting a single-sided electrode.

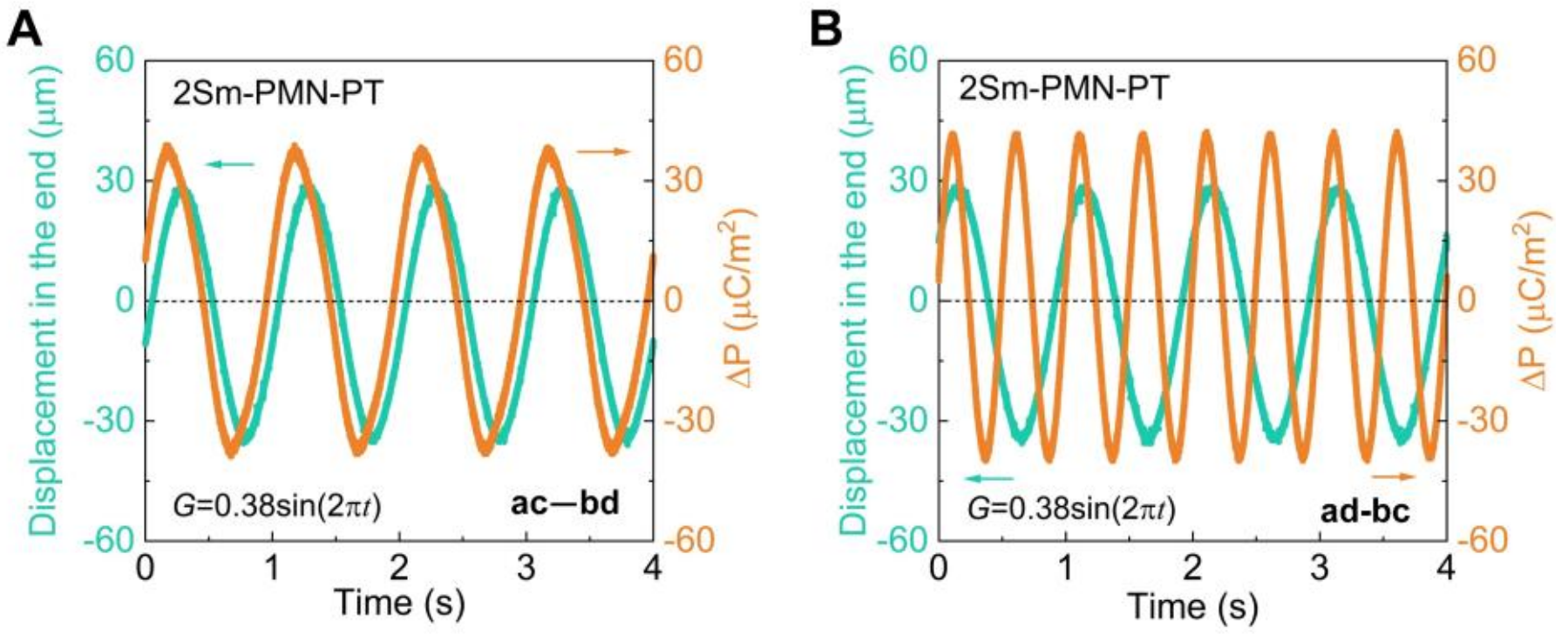

**Figure S17-2** (**A**) For linear effects in series ac-bd, the output signal has the same frequency as the displacement. (**B**) For high-order effects in series ad-bc, the frequency of the output signal is twice that of the displacement.

**Table S1**. The second order flexoelectric coefficients and *G*' for BTO slabs of different thicknesss.

| Thickness/mm | $w_{eff}$ / $\mu C$ | $G'/m^{-1}$ |
|---|---|---|
| 0.5 | 16 | 1.1 |
| 0.7 | 30 | 0.58 |
| 1 | 62 | 0.28 |
| 1.5 | 140 | 0.12 |